\documentclass[conference]{IEEEtran}

\usepackage[applemac]{inputenc}
\usepackage{amsmath}
\usepackage{amssymb}
\usepackage{epsfig}
\usepackage{hyperref}

\makeatletter
\newif\if@restonecol
\makeatother

\usepackage[boxruled,vlined]{algorithm2e}

\usepackage{graphicx}
\usepackage{adjustbox}

\usepackage{xcolor}
\usepackage{booktabs}
\usepackage{colortbl}
\usepackage{comment}

\usepackage[T1]{fontenc}

\usepackage{epstopdf}
\usepackage{array}
\newcolumntype{L}{>{\centering\arraybackslash}m{7cm}}
\newcolumntype{P}[1]{>{\centering\arraybackslash}p{#1}}
\newcolumntype{M}[1]{>{\centering\arraybackslash}m{#1}}
\usepackage[normalem]{ulem}
\usepackage{subfigure}

\definecolor{orange}{RGB}{255,127,0}

\renewcommand{\epsilon}{\varepsilon}

\newcommand{\Vis}{\textit{Vis}}

\newcommand{\layers}{\text{Layers}}
\newcommand{\filters}{\text{Filters}}
\newcommand{\grouping}{\text{Grouping}}

\providecommand{\DontPrintSemicolon}{\dontprintsemicolon}

\definecolor{tbllightblue}{rgb}{0.7294,0.8471,0.8980}
\ifCLASSINFOpdf

\else

\fi

\begin{document}
%
\title{DataSlicer: Task-Based Data Selection \\ For Visual Data Exploration}

\author{\IEEEauthorblockN{Farid Alborzi, Rada Chirkova,\\ Pallavi Deo, Christopher Healey,\\ Gargi Pingale, Vaira Selvakani}
\IEEEauthorblockA{North Carolina State University, USA\\
Email: \{falborz,rychirko,psdeo,\\healey,gpingal,vbselvak\}@ncsu.edu}
\and
\IEEEauthorblockN{Juan Reutter}
\IEEEauthorblockA{Pontificia Universidad Catolica de Chile\\
Email:  jreutter@ing.puc.cl }
\and
\IEEEauthorblockN{Surajit Chaudhuri}
\IEEEauthorblockA{Microsoft Research, USA\\
Email: surajitc@microsoft.com}
}



%


\maketitle

\begin{abstract} 
%
In visual exploration and analysis of data, determining how to select and transform the data for visualization is a challenge  
for 
data-unfamiliar or inexperienced users. 
Our main hypothesis is that for many data sets and common analysis tasks, 
there are relatively few ``data slices'' that 
result in effective visualizations. 
By  focusing human users on appropriate and suitably transformed parts of the underlying data sets, these data slices can  help the  users carry their task 
to correct completion. 

To verify this hypothesis, we develop a framework that permits us to capture exemplary data slices 
for a user 
task, 
and to explore and parse visual-exploration sequences  into a format that makes them distinct  and  easy to compare.  
We develop a recommendation system, DataSlicer, that matches a ``currently viewed'' data slice with the most promising ``next effective'' data slices for the given exploration task. We report the results of controlled experiments with an implementation of the DataSlicer system, using four common analytical task types. The experiments demonstrate statistically significant improvements in accuracy and exploration speed versus users without access to our system. 
\end{abstract}

\section{Introduction} 
\label{intro-sec}




Data-intensive systems accompanied by visualization software are being increasingly used for interactive data explorations \cite{StolperPG14,PolarisDiss,MackinlayHS07,parameswaran2013seedb,vartak2014seedb,
LivnyRBCDLMW97,StolteTH08}. These and other systems 
help data analysts  
in their exploratory tasks of visually identifying trends, patterns, and outliers of interest. 
The visualizations make it more efficient to find task-relevant types of objects in exploratory data analysis, especially in presence of large data. 
The reason is, visualizations allow analysts to leverage their visual pattern-matching skills, domain expertise, knowledge of 
context, and ability to manage ambiguity in ways that fully automated systems 
cannot. 

Due to the exploratory nature of their tasks, analysts often face a wide variety of visualization options to choose from. As pointed out in \cite{vartak2014seedb}, it is not the visualization per se that is the main challenge. Indeed, once the data to visualize have been selected and transformed (e.g., grouped and aggregated in an appropriate way), users can take advantage of a visualization tool to provide an effective visual presentation of the resulting data. In this paper we look into exploratory data analysis under the assumption that we have access to such {\em presentation} solutions, 
and focus instead on the issue of determining  which  ``data slices''  would be the most helpful to the user in addressing the task at hand when visualized. Here, the term {\em data slice} refers to the outcome of the process of selecting the data of interest from the given data set, as well as potentially  transforming  (e.g., grouping and aggregating) the selected data.  

Identifying the data slices that are appropriate for the given task 
is a challenge for inexperienced users or those not familiar with the data at hand. 
The reason is that, typically, only a small fraction of the available data slices results in task-relevant visualizations, while all the other options fails to help the user with her task. This may force such users 
to examine a large number of options, to find those that lead to relevant visualizations for their exploration or analysis task. While clearly a challenge in presence of large-scale data, this is a hard problem even when the data set is small.  



	


{\underline {\em Our Focus:}} 
Our focus is on analytical tasks of common interest, such as detection of outliers or trends, that users often perform in visual exploratory analysis of data.
Our objective is to improve the user experience by suggesting to her 
those data slices that, when visualized, present correct solutions to her task in a prominent way.  Solving this problem would be instrumental in helping casual or inexperienced users to effectively conduct 
explorations of potentially unfamiliar data sets,  in a number of application domains and for a spectrum of exploration objectives. 
 For our study, we assume that a user begins work by declaring the task that she plans to perform. We also assume that she is able to identify a correct solution for her task (e.g., an outlier) when the solution is presented to her prominently in a visualization of some data slice. 

\begin{figure*}[t]
  \centering
  \mbox{
    \subtable[A fragment of the data set \cite{datasetTask1Web} \label{task1-data-rows-fig}]
		{
			\begin{tabular}[b]{|p{1.5cm}|M{0.5cm}|M{0.5cm}|M{0.5cm}|}
    \hline
    \scriptsize{Place} & \scriptsize{AVG of Dep.} & \scriptsize{AVG of Mag.} & \scriptsize{NUM. of Rec.}\\
    \hline
    \scriptsize{Guadeloupe} & \scriptsize{100.0} & \scriptsize{7.4} & \scriptsize{1}\\
    \hline
    \scriptsize{Antigua and Barbuda} & \scriptsize{16.9} & \scriptsize{6.6} & \scriptsize{4}\\
    \hline
    \scriptsize{Martinique} & \scriptsize{102.0} & \scriptsize{7.0} & \scriptsize{3}\\
    \hline
    \scriptsize{East of Dominica} & \scriptsize{11.2} & \scriptsize{7.2} & \scriptsize{1}\\
    \hline
  \end{tabular}		
		}\quad
    \subfigure[A visualization using dimensions {\em average magnitude} (of earthquakes at location), {\em number of earthquakes} (at location), and {\em depth} (of earthquake) \label{raw-data-pic}]{\fbox{\includegraphics[width=5.8cm]{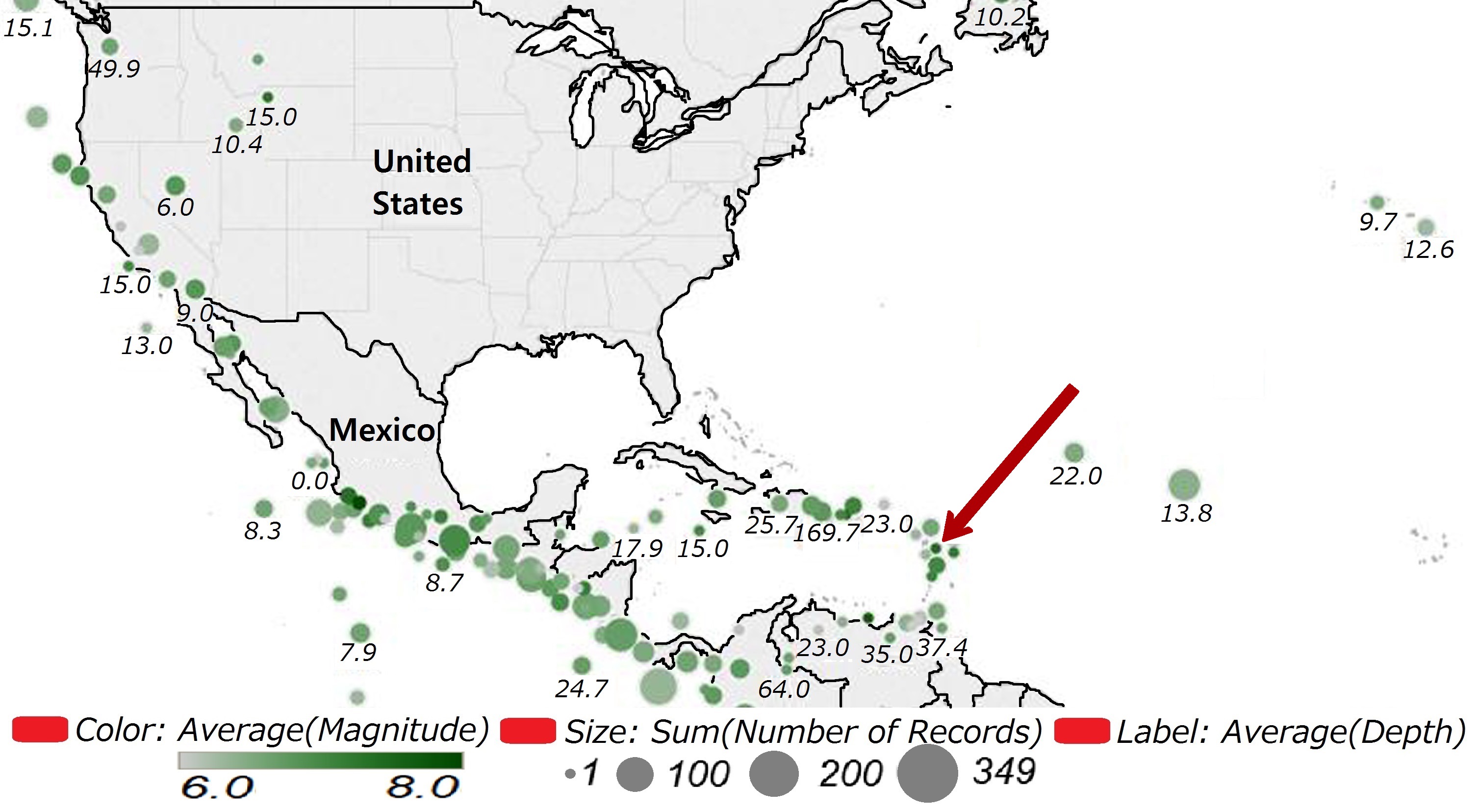}}}\quad
    \subfigure[A visualization using only the {\em average magnitude} dimension (bigger circles represent greater average magnitude) \label{dot-size-pic}]{\fbox{\includegraphics[width=5.8cm]{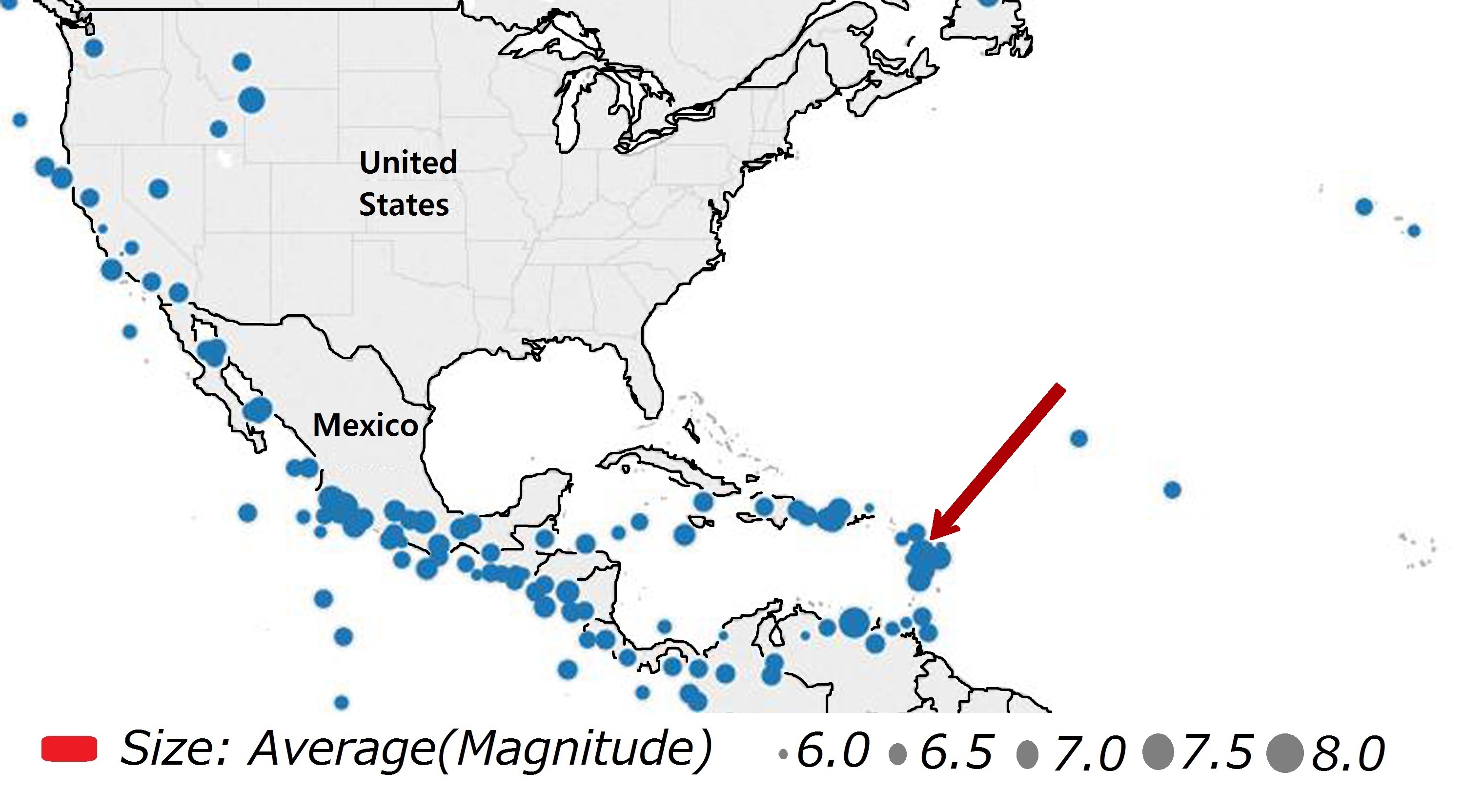}}}
}
\vspace{-5pt}
  \caption{Visual exploration (part 1) in search of earthquake-magnitude outliers in Central America using data set \cite{datasetTask1Web}, please see experimental task 1 in Section \ref{experim-results-sec}. The arrows in (b) and (c) highlight the visualizations of the ``Guadeloupe'' data point shown in (a); this data point is one of the answers to task 1.}

  \label{mainAfigureBlabel}
\vspace{-12pt}
\end{figure*}

\begin{figure*}
  \centering
  \mbox{
    \subfigure[A visualization using the {\em average magnitude} dimension (darker tones represent greater magnitude) \label{reduced-raw-data-pic}]{\fbox{\includegraphics[height=3.4cm]{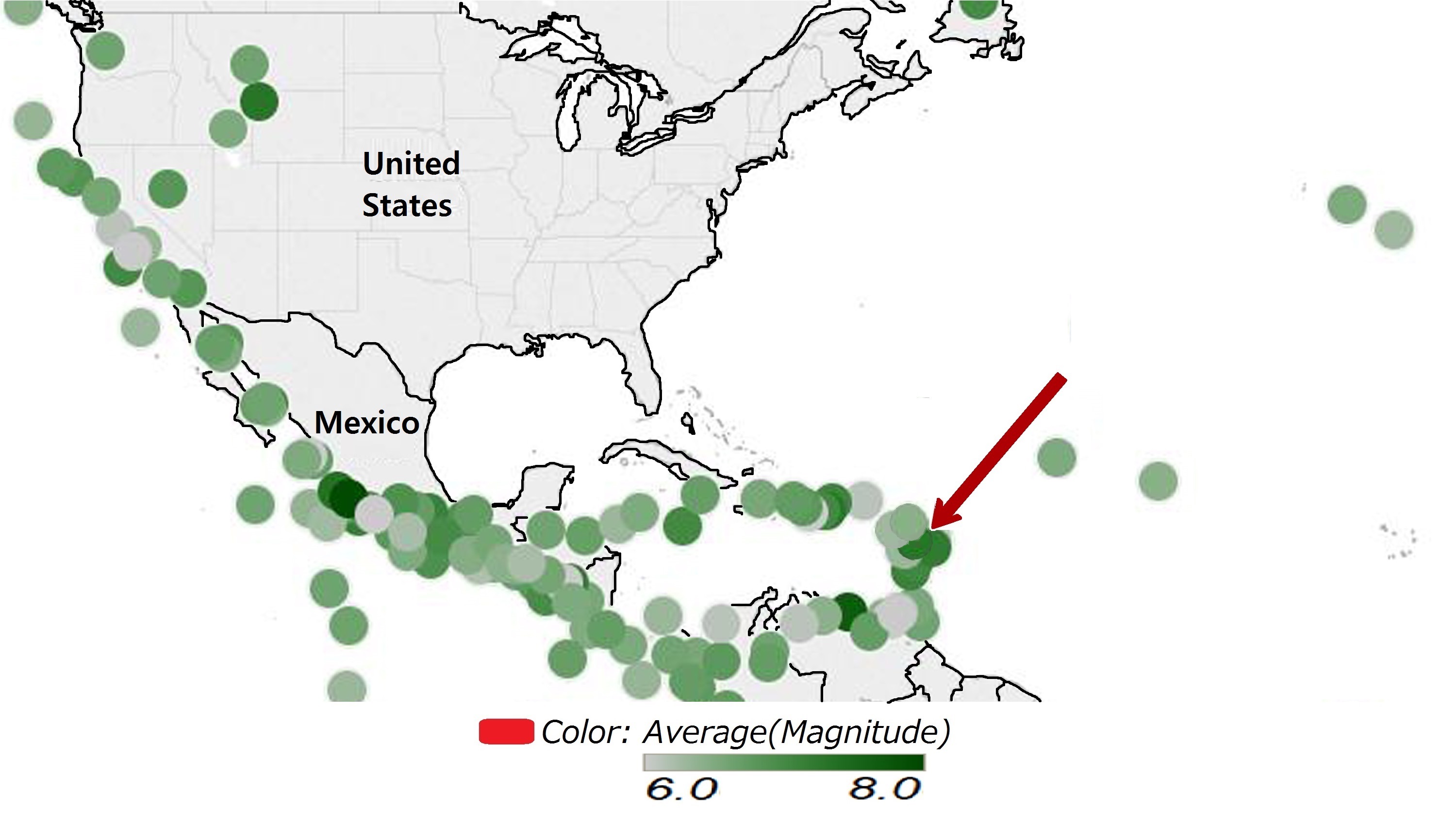
    }}}\quad
    \subfigure[Box plot showing outlier values of average earthquake magnitude\label{task1-boxplot-pic}]{\fbox{\includegraphics[height=3.4cm]{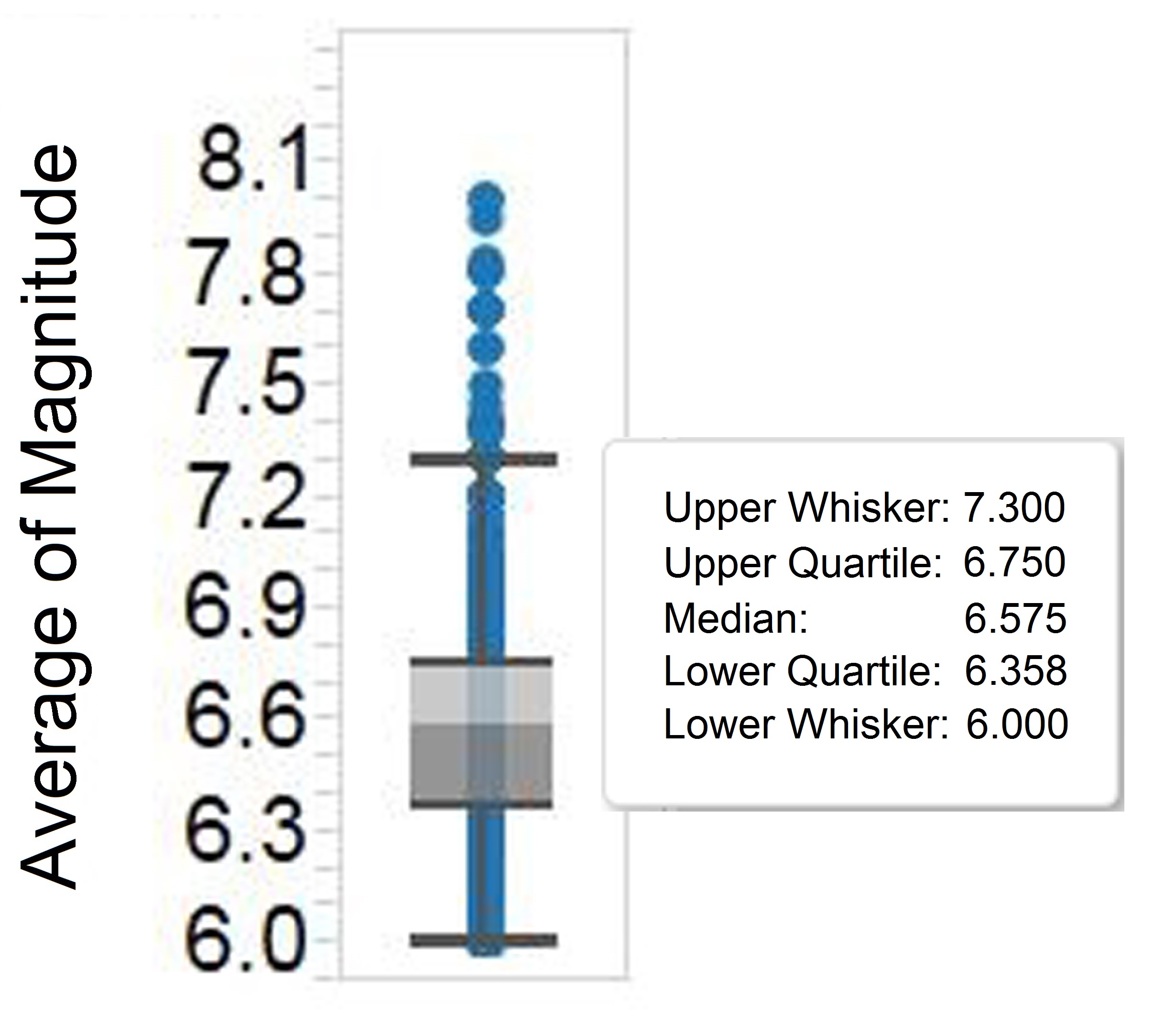}}}
    \subfigure[A visualization showing the answers (magnitude-outlier earthquake locations) prominently on the map\label{task1-secondstage-pic}]{\fbox{\includegraphics[height=3.4cm]{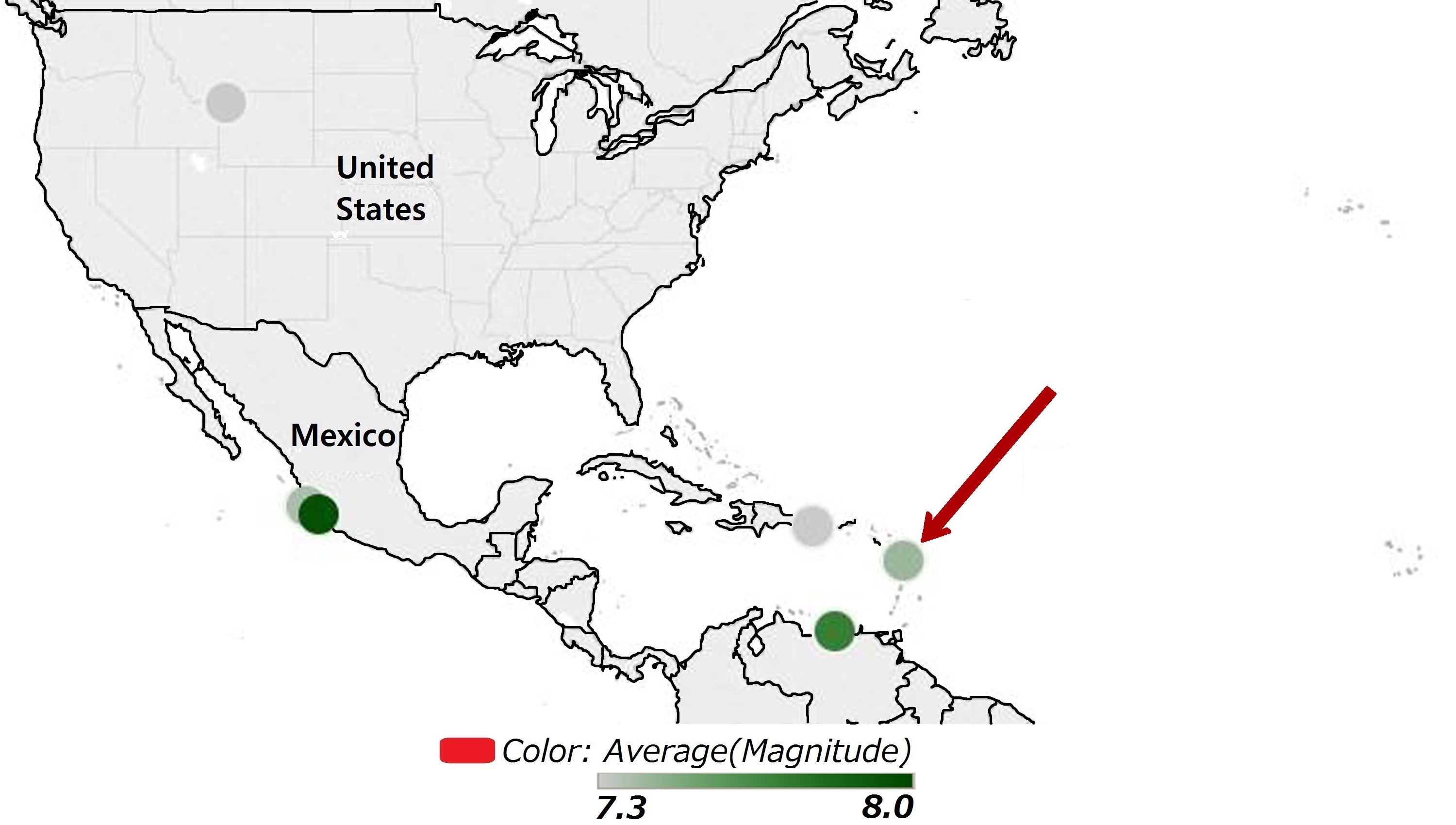}}}
}
\vspace{-5pt}
  \caption{Visual exploration (part 2) in search of earthquake magnitude outliers in Central America using data set \cite{datasetTask1Web}, please see experimental task 1 in Section \ref{experim-results-sec}.  The sequence (b)--(c) is an expert solution to the task. The arrows in (a) and (c) highlight the ``Guadeloupe'' answer data point, please see Fig. \ref{mainAfigureBlabel}.}
  \label{main figure label}
\vspace{-12pt}
\end{figure*}

{\underline {\em Proposed Solution:}} We address the combinatorial explosion in data-slice selection by basing data-slice suggestions on the stage at which the user is in solving 
her task, and 
(when available) on expert knowledge of the domain, task, and data set. In this emphasis on, and appreciation of, expert knowledge in solving complex data problems, our effort is in line with the research directions such as that of DeepDive  \cite{NiuZRS12,ShinWWSZR15}.



As an illustration, consider a relation storing the data from \cite{datasetTask1Web} (see \cite{jones2014communicating} for the details) 
 on major earthquakes worldwide from 1900--2013. The data set has 17 attributes and 8289 data points, please see Fig. \ref{task1-data-rows-fig} for a fragment of the data. 
 
 Suppose that in that data set, the {\underline {\em user task}} is to find locations in Central America containing earthquakes that are outliers based on magnitude.  In this user task, there is a wide range of options when selecting the initial data 
 to be visualized. 
 One natural starting point in the exploration 
would be to examine a map showing locations and other information about the earthquakes in the data set. 
One such visualization 
is shown in Fig. \ref{raw-data-pic}. 
The key point to note is that this visualization is unlikely to be helpful 
to those users who are not familiar with the data set. For instance, the arrow in Fig. \ref{raw-data-pic} is pointing to one correct answer (Guadeloupe in Fig. \ref{task1-data-rows-fig}) for this exploration task; observe that the visualization is not conducive to finding that answer, as the data point in question does not stand out in the visualization. 

One explanation for the relative ineffectiveness of the visualization of Fig. \ref{raw-data-pic} for the task at hand is that Fig. \ref{raw-data-pic} shows not only the location and magnitude, but also other information about each earthquake. 
Suppose 
%
%
the analyst eliminates those features of the data 
that are irrelevant to the task at hand; 
the resulting visualization could be as in Fig. \ref{dot-size-pic} or \ref{reduced-raw-data-pic}.\footnote{The difference between Figures \ref{dot-size-pic} and \ref{reduced-raw-data-pic} is just in the visual representation of the values of earthquake magnitude.} Interestingly and perhaps counterintuitively, we have found  that these visualizations are not very helpful either to human viewers performing this task on the data set \cite{datasetTask1Web}, 
again because the answers do not all stand out visually. 


%

A more effective way to address this exploratory task is for the user to first 
examine a box plot showing the earthquake-magnitude mean and outlier whiskers; please see Fig. \ref{task1-boxplot-pic} for the visualization. 
Once the cutoff value for outlier earthquake magnitude has been found, the user can effectively construct a correct answer for her 
task by 
filtering out the irrelevant data. The result is visualized in Fig. \ref{task1-secondstage-pic}. 


The data slice depicted 
in Fig.  \ref{task1-boxplot-pic} is not related to the data slices used to construct Figures \ref{raw-data-pic}--\ref{reduced-raw-data-pic}. The difference goes beyond removing irrelevant data features and, in fact, represents a drastically different choice of both the {\em data dimensions} and of their {\em grouping layout.} We found that if a user is unable to find a data slice that would present prominently the outlier values of earthquake magnitude in the data set, then suggesting to her the data slice (and the straightforward visual presentation) of Fig. \ref{task1-boxplot-pic} would typically enable her to proceed efficiently to constructing the data slice of Fig. \ref{task1-secondstage-pic}. Moreover, if the user is not sure how to proceed even after examining Fig. \ref{task1-boxplot-pic}, then she should find the data slice (and the map presentation) of Fig. \ref{task1-secondstage-pic} a helpful suggestion for the final stage of her overall task. 



In our experiments with this data set and user task (task 1 in Section \ref{experim-results-sec}), we found that for humans looking for earthquake-magnitude outliers  for the first time, it is not trivial to come up with an effective first-step visualization such as the box plot of Fig. \ref{task1-boxplot-pic}. Moreover, even though the data set \cite{datasetTask1Web} has relatively few (17) data attributes,  it is  impractical to enumerate all the possible data slices by brute force, in the hope of eventually identifying and visualizing a useful choice such as the data in Fig. \ref{task1-boxplot-pic}. Indeed, a seemingly natural but suboptimal choice of the initial visualization to look at --- such as those in Figures \ref{raw-data-pic}--\ref{reduced-raw-data-pic} --- is not necessarily conducive to finding the answers to the exploration task in question. While clearly a challenge in presence of large-scale data, this effect may be present even in those cases where the data sets are small by today's measures. (Recall that the earthquakes data set  \cite{datasetTask1Web} has 8289 records.) Note that relatively minor (``local'') modifications of initially suboptimal data choices to visualize, such as in the transition between Figures \ref{raw-data-pic}--\ref{dot-size-pic}, do not necessarily make the resulting visualization any more helpful to the user than the previous choice.

The {\underline {\em main hypothesis}} put forth in this paper is that for many data sets and common exploratory-analysis tasks, there are relatively few data slices that are key to providing effective visualizations for the task. Intuitively, these data slices are manifestations of the domain and data-set knowledge that is relevant to the task at hand. 
As we argue in this paper and corroborate with our preliminary experiments (see Section \ref{experim-results-sec}), the data-slice choices made by domain experts may help other users of the data set solve similar exploration/analysis tasks in a  more correct and efficient fashion. 
To substantiate and verify these claims, we use the specific measures (as in, e.g., \cite{HealeyE12,HealeyS12}) of:  {\em result accuracy,} 
understood as the average number of correct solutions
found, and of {\em user efficiency (speed),} 
understood as the average number of data-specification steps taken
to find a correct visualization for the task. 



Significant advances have been made lately in developing visual solutions for data exploration and analysis. Major projects, including  those described in \cite{dimitriadou2014explore,cetintemel2013query,drosou2013ymaldb,vartak2014seedb, parameswaran2013seedb,wasayqueriosity}, focus on determining which data slices could be useful to human viewers when visualized. (We provide an overview of these projects in Section \ref{rel-work-sec}.)   
Typically, data slices in these and other projects are suggested to the users based on generic expectations about what a user might find interesting in the data, rather than on the context of a particular task that the user might be facing, or of the user's stage in solving the task. Thus, to the best of our understanding, the solutions in the literature still fail to solve the problem of how to efficiently lead casual or inexperienced human users to visualizations of the data that summarize in an effective and prominent way the data points of interest 
for 
 the user's 
 exploratory-analysis task. As observed via the preliminary experiments reported in this paper, solving two distinct visual-exploration/analysis tasks on the {\em same} data set may lead to distinct sequences of data slices, with the data slices in each sequence being of value in the context, and perhaps at the specific stage, of just one of these tasks but not the other.  (Please see the discussion of experimental tasks 3 and 4 in Section \ref{experim-results-sec}.) In addition, to the best of our knowledge, suggesting (sequences of) data slices that would be helpful in solving at least one of these tasks, that of {\em determining trends in the data,} cannot be done using state-of-the-art tools. 

The specific contributions that we report are as follows: 

%
\smallskip
\noindent
$\bullet$ We develop a formal framework for capturing data slices of interest in a given class of visual-exploration tasks, and for providing appropriately visualized user-specific modifications of  each data slice. 
The data structures in the framework are scalable in the size of the data set, and typically do not need to be modified as the contents of the data set change over time. 
	
\smallskip

\noindent
$\bullet$  We develop prediction software that matches a ``currently viewed''  data slice with the most promising ``next effective'' data slice for the given type of exploration task on the data. 
	
\smallskip

\noindent
$\bullet$ We implement our framework and prediction system, DataSlicer, in tandem with commercial visualization software. 

\smallskip

\noindent
$\bullet$  Finally, we provide results from controlled experiments with 48 volunteers. The experiments demonstrate, for four common types of visual-analysis tasks, statistically significant improvements in accuracy and exploration speed versus users without access to our system. 
\smallskip

{\underline {\em Organization:}} After reviewing related work in Section \ref{rel-work-sec}, we present our framework in Section \ref{overview-sec}. 
Section \ref{data-viz-sec} outlines our main algorithms, and Section \ref{experts-sec} 
describes construction of our data structures. The architecture of the DataSlicer system is discussed in Section \ref{put-together-sec}. 
Section \ref{experim-results-sec} 
reports the experimental results, and Section \ref{sec-conlcusions} concludes. 

\section{Related Work} 
\label{rel-work-sec} 




Significant advances have been made lately in developing various facets of visual solutions for data exploration and analysis. 
In this space, we focus mainly on projects 
that concentrate on the problem of finding the right visualization, 
e.g., 
\cite{cetintemel2013query,parameswaran2013seedb,vartak2014seedb,wasayqueriosity}. 
We refer the reader to the survey \cite{idreos2015overview} for a more general discussion of data-exploration techniques. 

The system architecture in this current project is based on the connection between SQL queries and visualizations, which  
is at the core of commercial 
tools such as Tableau \cite{jones2014communicating,PolarisDiss}. Our data-slice format, as detailed in Section~\ref{data-slice-seq-sec}, has been inspired by, and is similar to, the formalization of visualizations provided in \cite{PolarisDiss}.  At the same time, the main purpose of that formalization in \cite{PolarisDiss} is for the visualization system to keep track of the current visualization, as it is being actively managed by the user, 
rather than by the system itself. In this current paper, the main purpose of the 
data-slice format is to match the user's current visualization with the stored past visualizations, and to recommend back to the user the best ``next-step'' data slice for her visualization sequence.  

As in \cite{GotzW09,GrammelTS10}, we view the task of constructing visualizations as a two-step process:  
One first decides on the data slice that is to be shown, and then chooses an appropriate visual 
specification for this data slice. Several projects, including \cite{GotzW09,GrammelTS10,HealeyD12}, have focused in this space on (semi) automatic recommendation of 
the best visual specification for a given task and data slice. However, the built-in assumption in those projects is that the appropriate data slice has been chosen. Our work is orthogonal to these efforts, in that we aim at choosing the best data slice, and assume that the visual specifications are given.  We expect to be able to combine forces in the future, to create a system that can help users to select both the appropriate 
data and the best presentation. 

Regarding the problem of choosing the appropriate data slice, the first connection that comes to mind is 
the problem of choosing the adequate SQL query for a 
given task. This problem has received substantial attention in the database community (see, e.g., \cite{fan2011interactive, khoussainova2010snipsuggest, chatzopoulou2009query}). At the same time, our work is more closely related to those projects that focus on learning which data need to be presented using a visual interface, rather than on constructing directly the appropriate SQL query. 
Here we have systems such as Vizdeck \cite{key2012vizdeck} and Charles \cite{sellam2013meet}, which aim to recommend the best 
visualization based on statistical properties of the data. There are also systems that 
recommend visualizations 
based on the user feedback 
\cite{cetintemel2013query,dimitriadou2014explore,drosou2013ymaldb}. The system called SeeDB 
\cite{vartak2014seedb, parameswaran2013seedb} 
automatically generates ``interesting 
visualizations'' based on those data slices where the trend deviates in a statistically significant way from the trend on the overall data set. Further, \cite{wasayqueriosity} describes a vision of an 
automated system, which 
can explore past user decisions  with the goal of discovering further operations on the data of potential interest to the same user. 

In this current project, our overall goal is the same as in the above papers. At the same time, instead of aiming for a fully automatic generic tool for selecting potentially popular individual data slices, we focus on choosing data slices 
that best address a given visualization-based {\em task.} As a result, the data slices selected by our system are task dependent, rather than just data-set dependent, and are also not limited to statistically interesting data. 
(For an illustration of how our system provides task-dependent, rather than data-dependent, recommendations, see Section~\ref{experim-results-sec} for experimental tasks 3 and 4 performed on the {\em same} data set.)  
Further, we work with the hypothesis that previous users, when faced with {\em the same type} of  task, could guide the system as to which data slices (or sequences thereof), with their visualizations, are interesting for the current user. In its emphasis on domain knowledge for the given task and data set, our approach is in line with research directions such as that of DeepDive \cite{NiuZRS12,ShinWWSZR15}.  
%
As a result, our approach can suggest to users data slices, such as those showing  general trends on the data, that state-of-the-art systems cannot recommend to the best of our knowledge. (See discussion of experimental task 4 in Section \ref{experim-results-sec}.) 


Finally, a good example of a collaborative tool for visualizing data 
is AstroShelf \cite{neophytou2012astroshelf}. This tool is specifically tailored for astrophysicists and, unlike ours, aims more at facilitating 
collaborations than recommending visualizations.


\section{The Framework: An Overview} 
\label{overview-sec} 


In this section we describe the envisioned user experience with a visualization-enabled system, 
where the system would advance the user's task-solving process by suggesting task-relevant data slices  from the underlying data. 
We then outline our proposed approach to delivering such an experience. 




\subsection{The Intended User Experience} 
\label{user-experience-sec}




When presented with a visual-exploration or visual-analysis task, users need to make decisions on which data  to visualize to solve the task. 
The default approach is for the user to construct various visualizations directly in a visualization tool, and to then keep improving or replacing them 
until one or more visualizations that are effective for the task are found. This can be time- and resource-consuming (cf. \cite{vartak2014seedb}).  Our goal is 
to alleviate or eliminate the inefficiencies in solving the data-selection part of the user's visual-analysis task. 

Our proposed system is designed to serve as a back-end of a standalone 
visualization tool. 
At any given time in working on the task, users 
may ask the system to suggest 
visualizations that would be useful for solving the task. 
If so requested, the system would analyze the current user's session and would recommend an (appropriately visualized) data slice based on the history 
of previous users who were involved in 
solving similar tasks. 
When analyzing the sequences of previous users, the system would assign higher priority to those data slices that were 
labeled by previous users as \emph{interesting}; 
for instance, a data slice is considered interesting if past users spent a considerable amount of time looking at its visualization(s).

Consider, for example, the task of finding earthquake-magnitude outliers in Central America using the data set \cite{datasetTask1Web}, as presented in Section~\ref{intro-sec}. 
A user may start her work on this task by constructing a visualization of Fig.  
\ref{raw-data-pic} or of Fig. \ref{reduced-raw-data-pic}. If she is overwhelmed by the amount of potentially relevant information in the visualizations, she 
would ask the system for a recommendation. The system would then analyze the user's current data slice, and would 
determine that the most successful past sequences involving the data slice of Fig. \ref{reduced-raw-data-pic} would  next 
switch to the data slice whose fragment is shown in Fig. \ref{task1-boxplot-pic}, and then to that whose fragment is shown in Fig. \ref{task1-secondstage-pic}. 
The two latter data slices, in that order and augmented by the current session's filtering conditions (Central America), would end up being chosen for the user. The system would determine appropriate visualizations for the recommended data slices by either using the user's visualization preferences in her current session or (if not available) by rules in the system. 
%
For the framework and system introduced in this paper, the claim of this example is corroborated by our experimental results, please see a discussion of experimental task 1 in Section \ref{experim-results-sec}. 






\subsection{The Proposed Approach: Data Sequences via Graphs} 
\label{main-overview-sec}

Our proposed framework and system are designed to work with users who create sequences of appropriately visualized data slices. A sequence could be exploratory, with the user 
trying to determine which individual (single) data slice works best for addressing her current task. Alternatively, a sequence could be part of a solution that calls for  construction of multiple consecutive 
data slices, as in the earthquake-magnitude task of 
Sections \ref{intro-sec} and \ref{user-experience-sec}. Either way, we use the graph representation to encode all the sequences of data slices for a  type of task on a data set;   we call 
the resulting graph the \emph{data-slice graph} for this task type and data set. In a data-slice graph, nodes encode data slices, together with any appropriate visualizations, and directed edges encode transitions between consecutive data slices in past user sessions. 

When users ask for recommendations, our system \emph{matches} their current session with the information stored in the data-slice graph, based on node similarity. Our approach can use any algorithm for measuring similarity between nodes; please see 
Section \ref{algorithms-sec} for a specific instantiation. 
The system then recommends to the user those data slices that were the most helpful, at the matched point in the graph, to previous users working on tasks of the same type. Again, our approach can use any algorithm for determining whether a node is helpful --- {\em interesting} --- enough to a user. (For instance, in our experiments we considered a data slice interesting if its visualization had been examined by at least one user for an amount of time above a fixed threshold.)
To enable the recommendation feature, each node in the data-slice graph is marked as either ``interesting'' or ``not interesting.'' 



The number of  data slices that one could construct using a data set with even a 
few attributes 
may 
be prohibitively large for computational purposes. 
It may not be practical or even feasible to represent and store all the possibilities explicitly. Instead, 
since our goal is to present the user with a specific data slice, 
we manipulate abstractions from visualizations 
using the relational model, similarly to what was done in \cite{PolarisDiss}.

More precisely, we map each data slice to a (simplified) relational-algebra expression, and work with relational queries. 
We store as nodes in a data-slice graph only those relational-algebra expressions that were featured in at least one  sequence executed for the same type of task on the data set at hand by at least one previous user. 

The data-slice graph contains all the information that we need to recommend data slices to the user: Once we match the 
user's current data slice to a node in the graph, it suffices to look for those interesting nodes in the graph that are 
``downstream closest'' to the matched node. Intuitively, this amounts to finding the next interesting nodes in previous sequences that feature a data slice 
similar to that of the current user. 
In the next two sections we provide details on 
the construction of the data-slice graph, how the matching is done, and how we look for the closest interesting nodes. 





\section{The DataSlicer System}
\label{data-viz-sec}


In this section we describe the DataSlicer framework and system. We start with a description of our theoretical framework 
for specifying sequences of data slices and their accompanying visualizations. We then discuss how the framework stores  
sequences in a data-slice graph, and explain how this graph is used to recommend to users 
data slices for addressing their task on the data set. 


\subsection{Data-Slice Sequences and Graphs} 
\label{data-slice-seq-sec}





We represent each visual depiction of data as a tuple $\Vis = \langle D, S \rangle$. Here, $D$ is the 
\emph{data specification}, which contains the information on the data slice in the visual depiction. 
Further,  
$S$ is the \emph{visual specification}, with information regarding how the data slice is to be visually presented, 
including the type of visualization (e.g., box plot or pie chart), colors, shapes, and so on. 
Consider, for instance, Fig. \ref{raw-data-pic}, which visualizes information on earthquakes in Central America. To create 
this visualization, we first need the latitude and longitude for each observation in the data set; this will tell us how to place each observation on the map. 
Fig. \ref{raw-data-pic} also shows three additional attributes for each observation point: the \emph{average magnitude}, the 
\emph{number of records,} and the \emph{average depth} of the earthquakes. Each attribute is shown using a different 
visual cue: We use the dot color to represent magnitude, the dot size to represent the number of records, and the dot label for the average earthquake depth. The visualization terminology for each of these attributes is a \emph{layer}; in general, each layer is 
assigned a different visual cue. 

Thus, the {\em data} specification $D$ for Fig. \ref{raw-data-pic} will state which information to extract 
about the 
data points to be shown: the latitude, longitude, magnitude, number of records, and depth, see Fig. \ref{fig-data-spec}.  
The  {\em visual} specification $S$ for Fig. \ref{raw-data-pic} states that the visualization needs to show the map of Central America, that each data point  
is to be shown as a dot, and what visual cue is assigned to each of the layers: color for average magnitude, size for number of records, 
and label for average depth. 


Our data-specification format has been inspired by, and is similar to,  the formal definition of visualizations provided in \cite{PolarisDiss}. 
(Please see Section~\ref{rel-work-sec} for a discussion of the difference between  \cite{PolarisDiss} and this project in the use of the formalism.) 
Similarly to \cite{PolarisDiss,vartak2014seedb}, 
we assume that the data to be specified come from a single relational table.\footnote{If two or more 
relations are to be visualized,  one could join them and treat the result as a single relation to be visually represented. This is a common approach in commercial 
data-visualization systems.} 
To define a data specification on a relation $R$, the following information is required:

\smallskip
\noindent
\hspace{10pt} $1$.
The fields applicable to the data set. These are either attributes of $R$ (called {\em simple fields),} or 
{\em complex fields} formed by combining two or more fields using the operations of concatenation ($+$), cross product ($\times$), and 
nesting ($/$) \cite{PolarisDiss}. 
We also allow aggregation over simple and/or complex fields, using operators \texttt{SUM}, \texttt{MIN}, \texttt{MAX}, or \texttt{AVG}.

\smallskip
\noindent
\hspace{10pt} $2$.
How the data from these fields are extracted. This amounts to specifying how the data are being grouped and which filters are currently active. Here we also provide information about which fields are being mapped to the visual axes {\tt X} and {\tt Y,}  
and about what fields are being rendered as layers. 
\smallskip


As an example of a data specification, consider again the visualization in Fig. \ref{raw-data-pic}. 
In this data specification, {\tt X} corresponds to longitude, {\tt Y} to latitude, and there are three layers:  \texttt{AVG}(magnitude), 
\texttt{SUM}(number of records) and \texttt{AVG}(depth). We also need to mention that the data are being grouped by 
the value of ``place.'' (The attribute ``place'' is a standard construct included in geographical data sets; it is used to group the data points by their geographical location.) The full data specification for Fig. \ref{raw-data-pic}  is shown in Fig. \ref{fig-data-spec}. 

\vspace{10pt}

\begin{figure}[t]
\begin{tabular}{| ll |}
\hline
simple fields: & lon (= longitude), lat (= latitude), \\
& pl (= place), mag (= magnitude), \\
&  nr (= number of records), de (= depth)  \\
complex fields: & $-$ \\
X Axis & lon \\
Y Axis & lat \\
$\layers$: & AVG (mag), SUM (nr), AVG (de) \\
$\grouping$: & pl  \\
$\filters$: & $-$ \\
\hline
\end{tabular}
\vspace{-5pt}
\caption{The data specification for the visualization of Fig. \ref{raw-data-pic}. }
\label{fig-data-spec}
\vspace{-10pt}
\end{figure}
\vspace{-10pt}

Formally, a {\em data specification} is a tuple {\bf $(X,$} $Y,$ $\layers,$ $\filters,$ $\grouping)$, where $X$ and $Y$ are 
the fields rendered respectively as the {\tt X} and {\tt Y} axis, $\layers$ is the set of fields rendered as layers, 
$\filters$ is the set of filters in use, and $\grouping$ is the set of attributes used for grouping. Continuing with our example, the data specification for Fig. \ref{raw-data-pic} is 

\noindent 
{\em (lon, lat, \{ \ AVG (mag), SUM (nr), AVG (de) \ \}, pl, - ).} 

A data specification is a SQL-query template of the form\footnote{This is the way specifications are generated in, e.g., the Polaris prototype \cite{PolarisDiss} of the Tableau software system \cite{jones2014communicating}.} 
{\small
\begin{verbatim}
SELECT <fields to be displayed> 
FROM <data set>
WHERE <filters on nonaggregated fields>
GROUP BY <grouping specification, X and Y axis>
HAVING <filters on aggregated fields>
\end{verbatim}
}

The connection between data specifications and SQL is important, as it provides flexibility when communicating 
with the log of visualization systems: We can either capture their data specifications, or we can capture SQL queries and produce specifications  
ourselves. For our example, the query is 

{\small
\begin{verbatim}
SELECT latitude, longitude, AVG(magnitude),
       SUM(number of records), AVG(depth)
FROM Earthquakes
WHERE latitude < 49.5 AND latitude > 5.3 AND
     longitude < -24.5 AND longitude > -128.7
GROUP BY place
\end{verbatim}
}



{\underline {\em The Navigation Algebra:}} 
We now specify operations on data specifications. The purpose is to enable transitions from one data specification to the next in a visual-exploration sequence that a user generates on the data. 
The basic operations for transforming data specifications are as follows: 

\smallskip
\noindent
$\bullet$ Add or remove a filter condition;

\noindent 
$\bullet$ Add or remove a field to/from the {\tt SELECT} condition (that is, the fields rendered as a layer), {\tt X} axis, or {\tt Y} axis;

\noindent 
$\bullet$ Add or remove a field to/from grouping specification; and

\noindent 
$\bullet$ Modify the specification of a complex field by adding or removing an operation (such as $\times$ or $+$). 

\smallskip
(In most systems, one can directly replace a field {\tt A} with a field {\tt B.} For technical reasons, we choose to model this action with two operations: 
removing {\tt A} and then adding {\tt B}.)  

We use the Navigation  Algebra to represent how users navigate between visualizations in a step-by-step fashion. 
Consider, for example, a user going from the visualization of Fig. \ref{raw-data-pic} to that of Fig. \ref{dot-size-pic}. 
We can model this as a sequence of three data specifications, starting with  

\noindent 
{\em (lon, lat, \{ \ AVG (mag), SUM (nr), AVG (de) \ \}, pl, - ),}

\noindent 
then removing depth, to obtain 

\noindent 
{\em (lon, lat, \{ \ AVG (mag), SUM (nr) \ \}, pl, - ),}  

\noindent 
and then removing the number of records, to arrive at 

\noindent 
{\em (lon, lat, \{ \ AVG (mag) \ \}, pl, - ),} 

\noindent 
which corresponds to the data specification of Fig. \ref{dot-size-pic}. 


{\underline {\em Sequences and Data-Slice Graphs:}} 
\begin{figure*}[t]
  \centering
  \mbox{
    \subfigure[\label{high-level-query-graph}]{\fbox{\includegraphics[trim=4 50 5 0,clip,width=8.0cm]{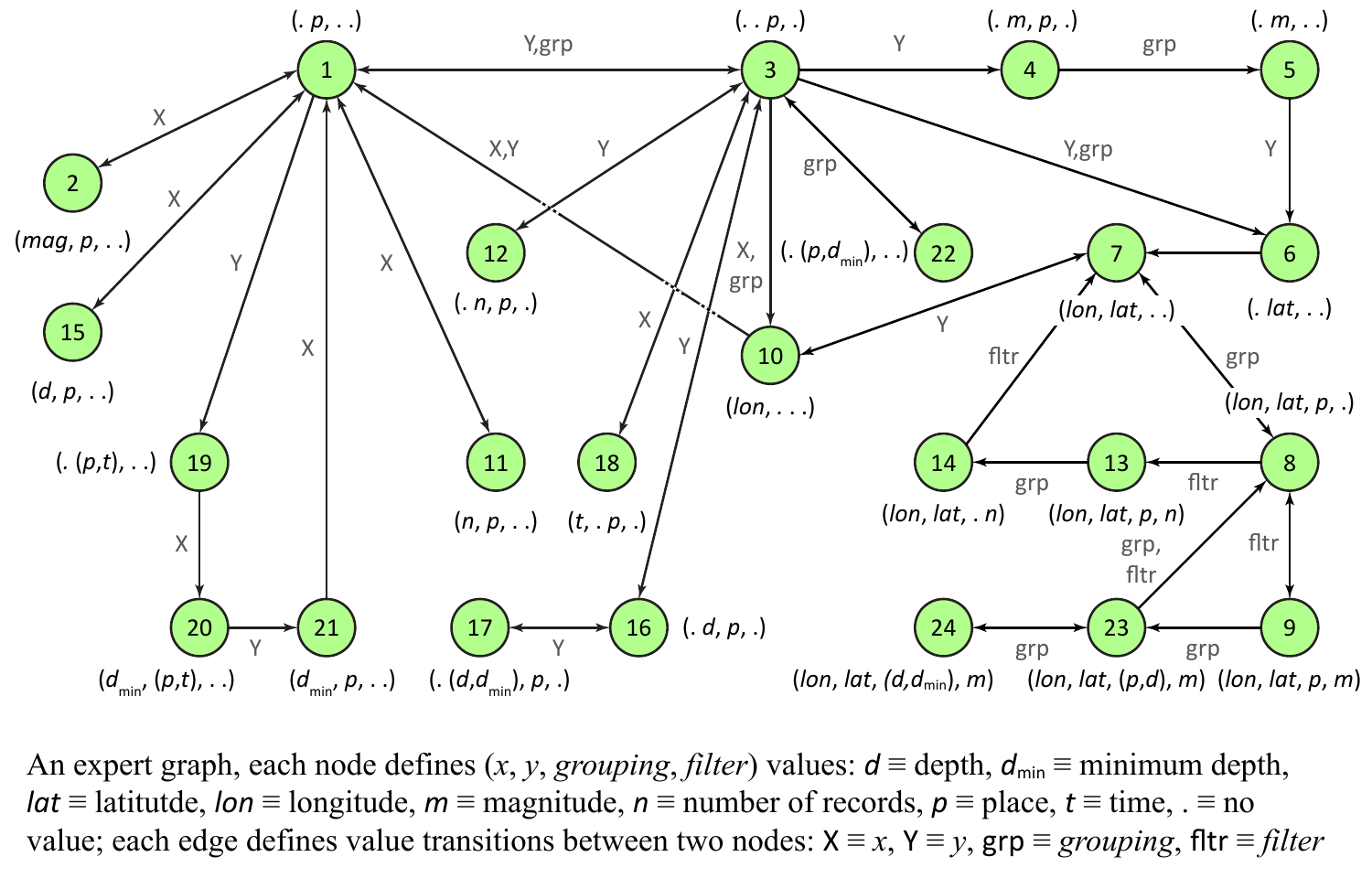}}}
    \subfigure[\label{zoomed-in-query-graph}]{\fbox{\includegraphics[trim=250 46 0 133,clip,width=8.0cm]{simplified}}}\quad
}
\vspace{-5pt}
  \caption{(a) A data-slice graph for experimental task 1 of Section \ref{experim-results-sec}. Each graph node is shown using its ($x$, $y$, $grouping$, $filter$) values, with $d=$ depth, $d_{min}=$ minimum depth, 
$lat=$ latitude, $lon=$ longitude, $m=$ magnitude,  $n=$ number of records, $p=$ place, $t=$ time, and . $=$ no
value. Edges define value transitions between nodes, with X $=$ $x$, Y $=$ $y$, grp $=$ $grouping$, fltr $=$ $filter$. In (b), the bottom-right fragment of the graph  
is shown at higher resolution.}
  \label{main figure label}
\vspace{-10pt}
\end{figure*}
When working on a visual-exploration or visual-analysis task, users create what we call sequences of visualizations: Starting at a 
particular visualization (such as that of Fig. \ref{raw-data-pic}), a user can 
create  new visualizations (such as the one of Fig. \ref{reduced-raw-data-pic}), by performing operations made available to them by the user interface --  
e.g., filtering the data, adding an extra attribute to the data specification, or changing the 
type of visualization. Each subsequent operation produces a new visualization in the sequence, and users continue 
in this fashion until  their task is complete. 

Our goal is to suggest to the user the slice of the data whose visualization is appropriate for the current stage of the user's task on the data. Thus, we do not concentrate on 
those parts of the sequences where new visualizations are created by modifying the visual specification. Rather, we  
focus on the underlying sequence given by the changes in the data specifications. 
These changes are modeled using our Navigation Algebra as described above. 
Assuming that we have a log with 
visualization sequences generated by previous users, 
we construct what we call the \emph{Data-Slice Graph} of this log: The nodes of this graph 
consist of all the data specifications occurring in the sequences in the log, and there is a directed edge from node $D_1$ to node 
$D_2$ if the log contains a sequence where $D_1$ and $D_2$ are consecutive data specifications. 

As an example, Fig. \ref{high-level-query-graph} shows a data-slice graph for the task of finding outlier earthquakes in the  
data set \cite{datasetTask1Web} (Section \ref{intro-sec}); this is task 1 in Section \ref{experim-results-sec}. The graph contains sequences generated by users 
who were solving the same type of task on the data set. Figure \ref{zoomed-in-query-graph} depicts a fragment of the graph, showing nodes with IDs $14$, $13$, $8$, $9$, $23$ and $24$. Figure \ref{zoomed-in-query-graph} was generated by the user sequence $(D_8$, $D_9$, $D_{23}$, $D_{24}$, $D_{23}$, $D_{8}$, $D_{13}$, $D_{14})$. 
The user started in node $8$, with the specification 
%
$D_8$ {\em = ( longitude, latitude, \{\}, place, - ),}  
%
that is, assigning the earthquake longitude to the {\tt X} axis, the latitude to the {\tt Y} axis, 
and grouping by \emph{place}. This specification corresponds to a visualization showing the map and just one 
dot for each place in that map where there has been at least one earthquake. (The grouping  in $D_8$ 
means that all the earthquake events in the same vicinity are grouped into a single tuple.)  
The  user then went on to add a filter on the attribute \emph{magnitude}, to filter out places where the average magnitude is not high enough. 
Note that rather than storing the precise filter, $D_8$ stores just the fact that a filter was added. This 
allows us to store together all the data specifications with 
similar filters. 
%
 
 Continuing with the sequence, the user then added depth (node ID $23$ with $D_{23}$) and minimum 
 depth ($24$ and $D_{24}$). Then the depth was removed, resulting in node $23$, and so on. 


{\underline {\em Interesting Nodes:}} 
Some of the sequences of visualizations in a log may contain data that are important for the user task.  
We denote these as \emph{interesting} visualizations, and mark these nodes as \emph{interesting nodes} in the data-slice graph.\footnote{In 
general, determining whether a visualization is interesting to a user is a nontrivial problem. While our framework can use any interestingness-measuring algorithm as a black box, 
in our experiments we marked as interesting all those visualizations which at least one user had visually examined for at least a fixed number of milliseconds.} 
For example, in our experiments with task 1 
of Section \ref{experim-results-sec}, the nodes with IDs $9$ 
and $23$   in 
Fig. \ref{zoomed-in-query-graph} were the most interesting to the human subjects. 
Since the data specification $D_9$ represents visualizations that are similar to that of Fig. \ref{task1-secondstage-pic}, this confirms the intuition 
that the visualization in Fig. \ref{task1-secondstage-pic} is amongst the most informative for this type of task. 

We distinguish between two types of users: {\em experts} and {\em regular users.} (This distinction is discussed in more detail in Section \ref{recommend-predict-sec}.) We say that there is an \emph{expert} (directed) edge 
from node $D_1$ to node $D_2$ if the sequence generating $D_1$ and $D_2$ was generated by an expert, and 
there is a \emph{user} (directed) edge if it was generated by a regular user. In addition, for each edge of the form $(D_1, D_2)$ we maintain 
with the edge the number of sequences in the log in which $D_2$ followed $D_1$.







\subsection{Algorithms to Match and Rank Data Slices} 
\label{algorithms-sec} 

The main focus of our framework is on servicing user requests to recommend the next task-relevant data slice and 
its appropriate visualization. To continue with our example, suppose that a user is exploring the earthquakes data set for 
magnitude outliers in Central America, and is currently looking at the visualization of Fig. \ref{raw-data-pic}. The data specification for Fig. \ref{raw-data-pic}, as 
discussed in Section \ref{data-slice-seq-sec}, is 

\noindent 
{\em (lon, lat, \{ \ AVG (mag), SUM (nr), AVG (de) \ \}, pl, - ). } 
 
\noindent 
When the user asks for a recommendation, the system needs to perform the following two operations: 


\smallskip
\noindent
\hspace{10pt} $1$.  
The data specification currently being examined by the user needs to be matched to data-specification nodes in the data-slice graph. 
We keep all such ``best-match'' nodes. 
 
\smallskip
\noindent
\hspace{10pt} $2$. 
 Once a match has been found, 
the system needs to find in the data-slice  graph those ``downstream'' data specifications  
that are potentially interesting to the user 
and are at the same time the closest to the matched node, in terms of operations of the Navigation Algebra. 
\smallskip


The algorithm addressing the first challenge is called {\tt Match Data Slices}, please see  
Algorithm \ref{algo:Match-function} for the pseudocode. The algorithm accepts a data specification $D$ and computes, for the stored data-slice graph $G$, the edit distance between $D$ and the data specification in each node of  
$G$. (As mentioned in Section \ref{overview-sec}, both this algorithm and the {\tt Rank Data Slices} algorithm can use any distance measure, e.g., page rank. The edit distance shown in the pseudocode of {\tt Match Data Slices} is one specific choice made in our implementation described in Section \ref{put-together-sec}.) We do not want to differentiate between the specifications where the X and the Y  axis are switched, as they represent 
semantically the same object, and likewise for switching between layers and axes. Thus, we proceed as follows. 
For each node $n$ in the graph we compute three distances between $n$ and $D$: (1) The edit distance $d_s$ that considers only 
the fields assigned to the $X$ and $Y$ axes and the layers in $D$ and $n$; (2) the edit distance $d_g$ considering only the fields in the 
grouping clause; and (3) the edit distance $d_f$ that considers only the filters in each of $D$ and $n$. We then add the three values, and output all the nodes $n$ in the graph for which the resulting value is the lowest.


 
We now look at addressing the second challenge listed above, making recommendations using the current match. 
Once we have matched a specification to a node in the data-slice graph, the next task is to retrieve the interesting 
``downstream'' nodes in the graph that 
are the closest to the matched node. We do this using our {\tt Rank Data Slices} algorithm, please see  
Algorithm \ref{algo:Ranking-Algorithm} for the pseudocode. The algorithm works as follows. We assume that each node $k$ in the data-slice graph 
is given an ``interestingness'' value $I_k$. (Any interestingness measure will work for our purposes, as outlined in Section \ref{overview-sec}.) We are also given a threshold $T$, with the objective of  
selecting only those nodes with an interestingness value above $T$, as well as the desired number $M$ of output nodes. 
For each node $n$ that is in the output of {\tt Match Data Slices,} we select all the nodes in the data-slice graph whose interestingness 
value is greater than the threshold $T$, and rank them in terms of their weighted-shortest-path distance to $n$. (Other distance measures could be used instead.) We then select and return the $M$ nodes 
from this list that are closest to $n$; if there are not enough such nodes, we complete the list with the most interesting nodes overall according to the 
 $I$-values in the graph. (This might be necessary if, for instance, the user's current visualization is not relevant to the task and thus cannot provide a useful input to the {\tt Match} algorithm.)

In our experiments, 
as reported in Section~\ref{experim-results-sec}, we chose screen time as our measure of interestingness of each data specification. (We assume there that the longer a user looks at the screen in examining a particular visualization, the more interesting that 
visualization is to the user.) We also set our threshold $T$ to 3 seconds. 
Though it might look like a small value for the interestingness threshold, its effect is that of filtering out almost 70\% of the graph nodes. 
Furthermore, in the experiments we considered the graph information that had originated 
from an expert as more helpful than the information from a regular user, and thus made the weight of expert edges in the graph {\em lower} (i.e., intuitively contributing to a shorter distance from the matched node) than the weight of ``regular-user'' edges. Specifically, the weight of an edge from a specification $D$ to a specification $D'$ 
that was part of an expert sequence 
would be set in the experiments to $1$, and the weight of an edge from a regular-user sequence would be set to $1 + 1/n_u$, 
where $n_u$ is the total count of previous users' sequences that have moved from $D$ to $D'$ in one step. Please see Section \ref{recommend-predict-sec} for a discussion of expert and regular edges. 



Coming back to our example, recall that the specification of Fig. \ref{raw-data-pic} was matched to the nodes $8$ and $23$ of the query 
graph. A call to {\tt Rank Data Slices} will now try to find the most interesting specifications that are closest to these nodes. 
Intuitively, this can be understood as asking for the most interesting specifications that include the latitude and longitude,  
and thus are expected to be shown in a geographical representation. 
The ranking algorithm would return the two interesting nodes that are closest to either $8$ or $23$; these answers include $23$ itself, with distance $0$, and $9$, with distance $1$.  To present these back to the user, we take these specifications and produce a visualization using the user's previous visual specification, which was a geographical representation. If we use the visual specification of Fig. \ref{raw-data-pic}, the visualization of the 
specification of node $9$ would look  like that of Fig. \ref{task1-secondstage-pic}.  








\begin{algorithm}[t]
\DontPrintSemicolon
\textbf{Given: } Data-slice graph $G$ and maximal number $M$ of nodes in the output.

\KwIn{Data specification $D$.}
\KwOut{The set of nodes of the data-slice graph $G$ that is nearest to $D$ in terms of edit distance.}
 
\For{each node $n$ in $G$} {
	$d_s=$ edit distance between the selection arguments of $n$ and $D$;
	
	$d_a=$ edit distance between  the filters of $n$ and $D$;
	
	$d_g=$ edit distance between the grouping clauses of $n$ and $D$;
	
	set $distance(n) = d_s + d_a + d_g$;
}
  
\Return{up to $M$ nodes $n$ in $G$ with the lowest values of $distance(n)$.}
\caption{Match Data Slices}
\label{algo:Match-function}
\end{algorithm}

\begin{algorithm}[t]
\DontPrintSemicolon
\textbf{Given: } Data-slice graph $G$, maximal number of output nodes $M$, interestingness value $I_k$ for each node $k$ in $G$, and interestingness threshold $T$.

\KwIn{A node $n$ of the data-slice graph $G$.}
\KwOut{List of $M$ interesting nodes closest to $n$.}
$L = \emptyset$;

\For{each node $m$ in $G$ with $I_m$ $>$ $T$} {
	compute $distance(m) = weighted\_shortest\_path(n,m)$;
	
	\If{$distance(m) < infinity$} {
		$add\textit{ }m\textit{ }to$ $L$;\;
	}
}

Sort $L$ according to distance (ascending);

\If{$ size(L) >= M$} {
		return the $M$ first nodes of $L$;\;
	}
\Else{add to $L$ the $M-size(L)$ most interesting nodes in $G$ according to $I$ that are not in $L$;}
\Return{$L$.}
\caption{Rank Data Slices}
\label{algo:Ranking-Algorithm}
\end{algorithm}

\section{Constructing and Using Data-Slice Graphs} 
\label{experts-sec} 

In this section we outline the process of 
constructing the data-slice graph for a given task type on a data set. Then we discuss the modes of using data-slice graphs depending on whether domain experts have been involved in the construction. 

\subsection{The Construction Algorithm} 
\label{construc-sec}

Recall (Section \ref{intro-sec}) that we assume that each user declares her task as she begins the work. Thus, each user sequence can be associated in the log 
with the task that the user was solving when generating the sequence. We also assume that each expert sequence (if any) is marked as such by the log administrator; we discuss the implications later in this section. At the point of logging a completed user sequence, we reformulate it, with two goals in mind. 
First, we make sure that all the logged sequences 
are formulated ``at the same level of granularity.'' Toward this goal, we 
make each sequence detailed enough so that each edge in the output sequence corresponds to a single operation in the Navigation Algebra of Section \ref{algorithms-sec} (see Fig. \ref{high-level-query-graph} for an illustration of the outcome). The second goal is to mark, in each sequence, each  node that is interesting under the given interestingness measure, see Sections \ref{main-overview-sec} through \ref{algorithms-sec} for a discussion. The overall algorithm for this reformulation of user sequences is straightforward. 

Suppose now that we have selected from the log all the sequences that are to be included in the data-slice graph that we are constructing. (We discuss potential selection criteria in Section \ref{recommend-predict-sec}.) We begin the construction by declaring one arbitrary selected sequence as the (initial) data-slice graph. We then keep adding the other selected sequences to the graph one at a time, by combining each node in the current sequence with some node in the graph, as long as the two nodes are the same in the $D$ part of their $\Vis = \langle D, S \rangle$ representation. That is, we combine a node in a user sequence with a node in the graph if and only if the $D$ parts of these nodes are the same; we store with each resulting node as many visual ($S$) specifications as we had in all the 
nodes that we have combined. If, on the other hand, for a node $n$ in the sequence being added there are no nodes in the data-slice graph that have the same $D$ part as $n$, we just add $n$ as a new node in the graph. For each node we keep the maximum interestingness amongst all the sequences in which this node appeared. Once we have merged all the nodes of a sequence with the graph in this manner, we add to the graph all the edges belonging to the sequence being added. In the process, if the sequence being added is an expert sequence, we re-weigh its edges as described in the discussion of the {\tt Rank Data Slices} algorithm in Section \ref{algorithms-sec}. 


{\em Output and Correctness:} The output of the overall graph-construction algorithm is a data-slice graph constructed as described above. By definition of  the algorithm, its output does not depend on the order in which the selected input sequences are processed and merged with the graph. The construction can be done either in the batch fashion  or with the graph being enhanced over time in an incremental fashion, with addition of one user sequence at a time as needed. 

\subsection{Recommendation and Prediction Systems} 
\label{recommend-predict-sec}

We now discuss possible criteria for selecting logged user sequences for entry into the data-slice graph. 

{\underline {\em Recommendation Systems:}}  
One criterion could be to include all the sequences from the log that are associated with the task type of interest. (We consider two tasks on the same data set to be of the same type if they differ only in the filtering criteria. E.g., we declare to be of the same type the tasks ``find all the magnitude-outlier earthquakes in the world'' and ``find all the magnitude-outlier earthquakes in Central America'' on the data set \cite{datasetTask1Web}, see Section \ref{user-experience-sec}.) In this case, there is no need to mark user sequences as expert, and thus the entire process of constructing both the log and the data-slice graph as described in Section \ref{construc-sec} can  be fully automatic.

We call such a data-slice graph a {\em recommendation graph;} the overall DataSlicer system will function as a recommendation system in this case. The reason is, in this case we have no information on which nodes in the graph would be the most helpful to the users in prominently featuring  correct solutions to their task. In working with such a graph, the users will possibly ``upvote'' over time those graph nodes that are more helpful to them in solving their task. This ``upvoting'' process is sound, as we assume (Section \ref{intro-sec}) that each user can recognize correct solutions once they have been presented to her prominently in some visualization of the data. (The ``upvoting'' functionality can be easily added to the ranking algorithm of Section \ref{algorithms-sec}.) The resulting graph nodes can then be recalibrated automatically into more interesting nodes. 

{\underline {\em Prediction Systems:}}  
We now consider the case where the help of domain experts is available, or perhaps even sought after, as would be in case of mission-critical applications. Recall that the log administrator can mark some of the sequences to be logged as coming from domain experts. This can be done in case one or more experts on the domain, task type, and data set are involved in solving tasks of this type for the benefit of the user community; the community could be employees of a certain company, analysts using a certain product, and so on. In this case, the process of constructing the graph is the same as before (see Section \ref{construc-sec}), with expert nodes and edges being marked explicitly as such in the construction. 

When the DataSlicer system uses a data-slice graph constructed using expert sequences, we refer to this mode of operation as ``prediction mode,'' and to the system as a ``prediction system.'' Indeed, domain experts are expected to know how to solve effectively and efficiently tasks of the given type on the data set, and nodes and edges generated in the 
graph by their solutions are expected to help the community in solving tasks of the same type more so than sequences created by regular users. Note how our algorithm of Section \ref{construc-sec} incorporates into a data-slice graph and automatically reconciles potentially different approaches of multiple experts to solving the same task. As a result, sequences coming from multiple experts get transformed into multiple solution paths in the graph. 

\section{Putting It All Together} 
\label{put-together-sec} 



\begin{figure}
\centering
\includegraphics[width=9cm]{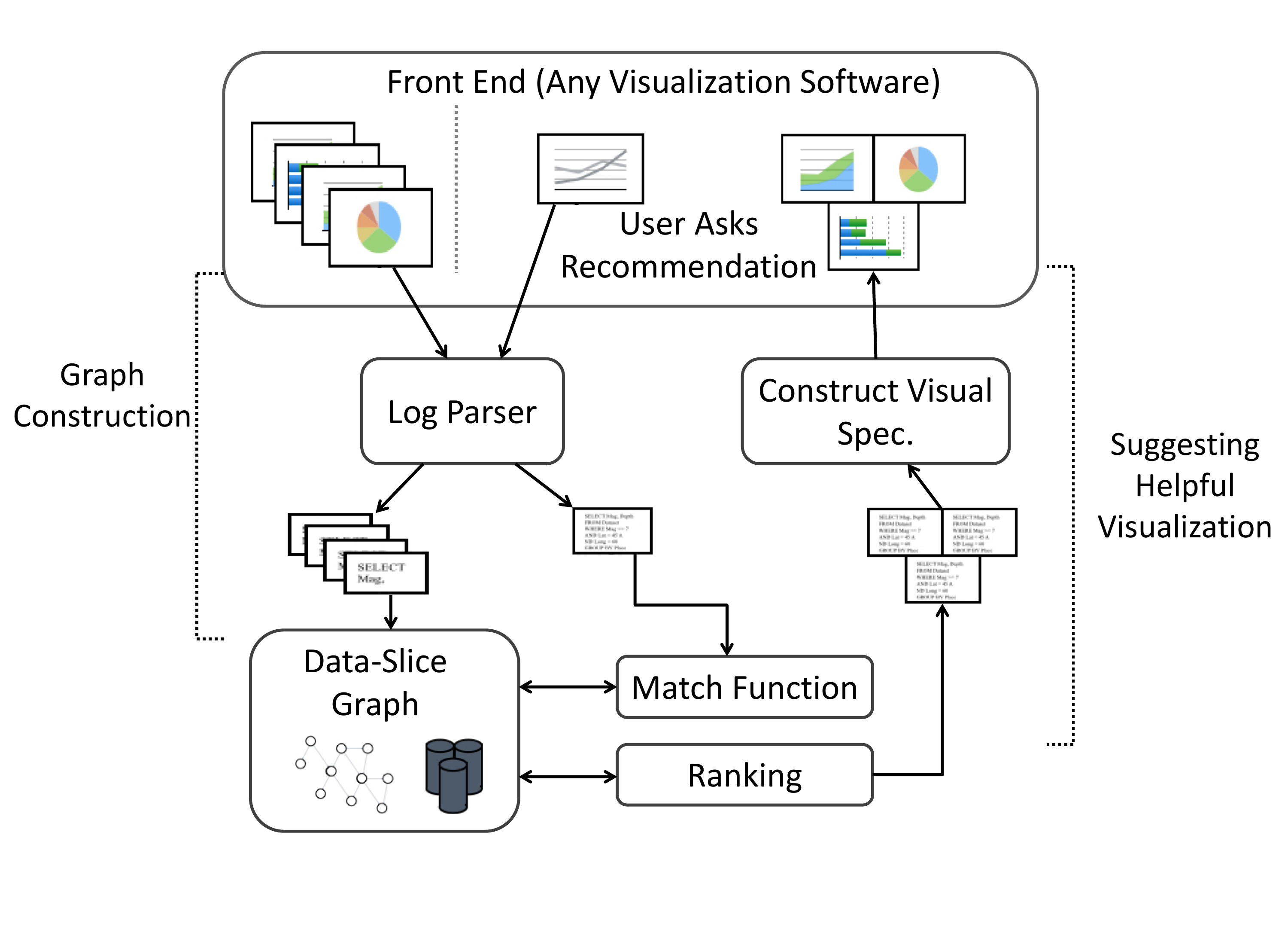} 
\vspace{-35pt}
\caption{The DataSlicer system architecture.}
\vspace{-5pt}
\label{architecture-pic}
\vspace{-10pt}
\end{figure}

Fig. \ref{architecture-pic} depicts a high-level overview of the architecture of our system. In this section we outline it 
component by component, and then discuss the scalability and implementation. 

{\underline {\em Front End:}} The front end of the system can be any visualization tool, as long as it can issue appropriate data-specification queries on the data-set store and visualize the answers, and also has a means of communicating 
its operations to other software. Some commercial visualization systems make available logs of their operations; 
we have implemented DataSlicer with a commercially available front-end tool, in such a way that all the DataSlicer communication 
with the front end is done through such logs, as explained in the next paragraph. 

{\underline {\em Interface:}} The DataSlicer interface is the means to connect with the front-end visualization tool. The interface is in charge of the following 
two main tasks: First, it provides a way to obtain and understand logs of the system, to enable extracting from the logs 
information about previous-user  sessions. This part of the interface is called the \emph{log parser}; it also maintains the current user's current 
visual specification, as well as the data specifications returned by the ranking algorithm of Section \ref{algorithms-sec}.
Second, once the system has recommended a set of data specifications, the interface visualizes them and presents them back to the user. 
To create these visualizations, we maintain the current user's previous visualization preferences and use them wherever possible  to visualize the recommended data specifications. For  those recommended data specifications that cannot be visualized using the current user's visual preferences, the system  
uses default visualization rules. 
Because of the closed architecture that many commercial visualization systems opt to implement, for our experiments 
we had to implement this second task in a semi-manual way. 

{\underline {\em Data-Slice Graph:}} The data-slice graph for the given task type and data set is physically stored as a separate database.
 We do not allow for any direct updates to the data-slice graph. 
Instead, to augment the data-slice graph with more information we set up separate system sessions, where past users sequences are provided to the log parser. During those sessions, the log parser 
enhances the existing data-slice graph with a new set of sequences, or creates a new data-slice graph from scratch, as 
detailed in Section \ref{construc-sec}. 

{\underline {\em Back End:}} The back end of the system is the part that is in charge 
of producing recommendations for users. It comprises the Match and Rank algorithms, as described in Section \ref{algorithms-sec}. 

{\underline {\em Scalability:}} In the DataSlicer architecture, visualizations are constructed for users by separate front-end visualization software, which sends to the data store queries based on the data slices, and then visually postprocesses the query answers. Thus, in the overall  DataSlicer system, the processing of data-slice queries is decoupled from executing the Match and Rank algorithms of Section \ref{algorithms-sec} on the data-slice graphs. Further, data-slice graphs are constructed based on task-exploration sequences, and thus on the structure rather than on the contents of the data set being explored. Thus, the size of a data-slice graph does not depend on the number of tuples of the data set, 
and the graph does not need to be modified as the contents of the data set change over time. 
On the other hand, the size of a data-slice graph is directly proportional to the number of user sequences that it captures, 
and the Match and Rank algorithms clearly  run in at most linear time with respect to the size of the graph. 
Addressing the issue of scalability of Match and Rank in the number of user sequences in the data-slice graph is a direction of our current work.

{\underline {\em The Implementation for the Experiments:}}  
The system used for the experiments reported in Section \ref{experim-results-sec} has been built using the Java framework and compiled using JDK 1.8. To store the data-slice graphs for the experiments,  we used MongoDB version 2.2. 
We worked with a commercial visualization tool; we can support working with any visualization tool, but for each different 
visualization tool, a different DataSlicer interface needs to be built. (This includes the log parser and the connection that  
presents visualizations back to the user.)  

\section{Experimental Results} 
\label{experim-results-sec}

\begin{figure*}
  \centering
    \subtable[Fragment of data set \cite{datasetTask2Web} for task 2\label{task2-data-rows-fig}]
		{\raisebox{3mm}{
	\resizebox{5.5cm}{!}{%
		\begin{tabular}[b]{|p{2cm}|M{1.0cm}|M{0.5cm}|M{0.5cm}|}
    \hline
    \small{Name} & \small{Position} & \small{Age} & \small{BA}\\
    \hline
    \small{Melky Mesa} & \small{UT} & \small{25} & \small{0.50}\\
    \hline
    \small{Derek Jeter} & \small{SS} & \small{38} & \small{0.32}\\
    \hline
    \small{Andy Pettitte} & \small{P} & \small{40} & \small{0.25}\\
    \hline
    \small{Francisco Cervelli} & \small{C} & \small{26} & \small{0.00}\\
    \hline
    \small{Chris Dickerson} & \small{OF} & \small{30} & \small{0.29}\\
    \hline
    \small{Brett Gardner} & \small{LF} & \small{28} & \small{0.32}\\
    \hline
    \small{Rabinson Cano} & \small{2B} & \small{29} & \small{0.31}\\
    \hline
    \small{Eric Chavez} & \small{DH} & \small{34} & \small{0.28}\\
    \hline
	  \end{tabular}%
  	}
		}}\quad
    \subfigure[Box plot showing prominently answers (outliers) for task 2 on data set \cite{datasetTask2Web} \label{task2-outputs-pic}]{\includegraphics[width=9.5cm]{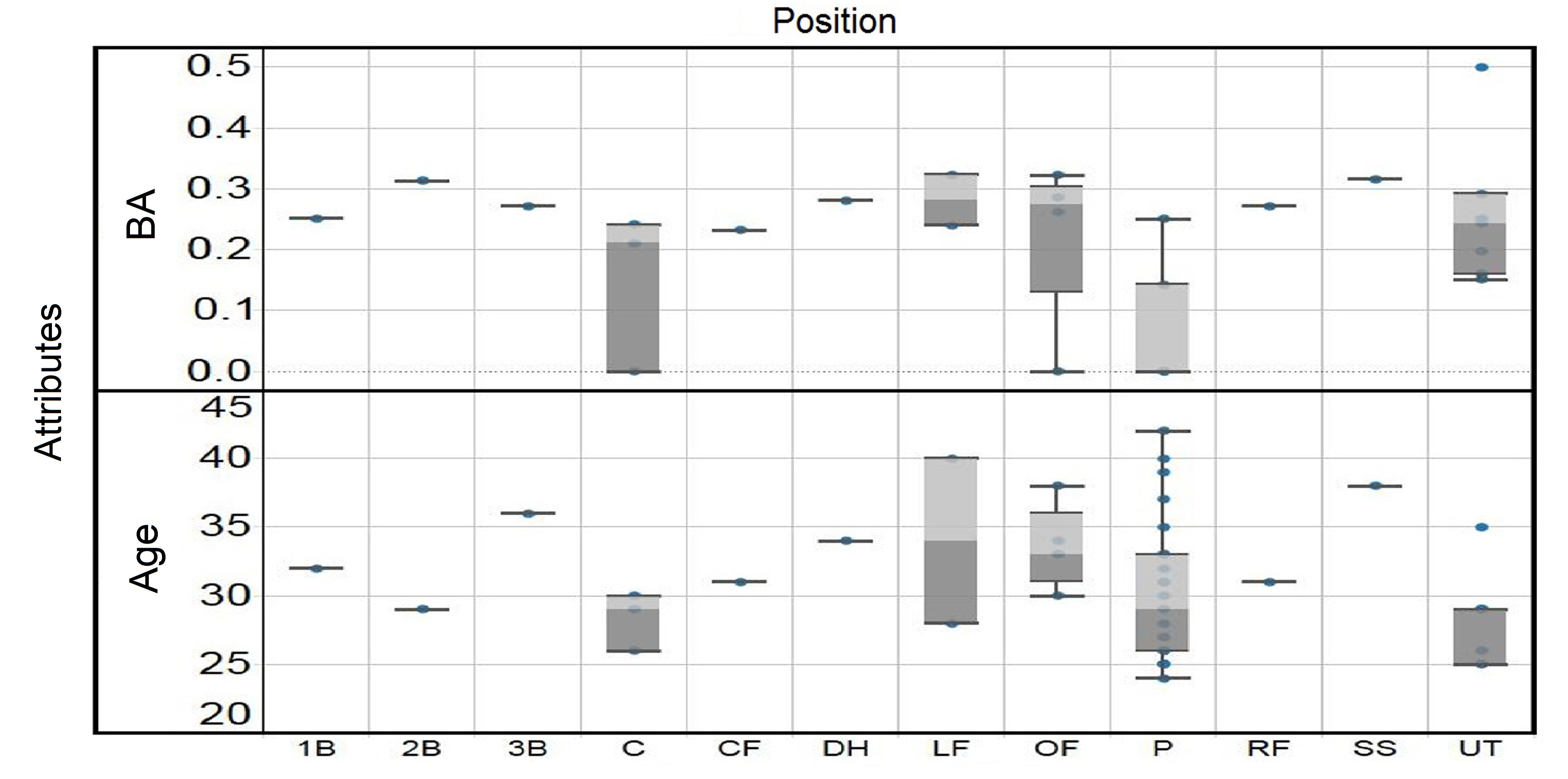}}\quad
    \vspace{-5pt}
  \caption{Experimental task 2 using data set \cite{datasetTask2Web}: fragment of the data set and a visual solution that shows prominently the answers to the task.}
      \vspace{-5pt}
  \label{main figure label}
\end{figure*}

\begin{figure*}
  \centering
  \mbox{
    \subtable[Import costs (\$ / container) in 2005--2010 in \cite{datasetTask3Web} \label{task3-data-rows-fig}]
	{\raisebox{7mm}{
		\resizebox{0.8\columnwidth}{!}{%
		\begin{tabular}[b]{|p{1.9cm}|M{0.6cm}|M{0.6cm}|M{0.6cm}|M{0.6cm}|M{0.6cm}|M{0.6cm}|}
    \hline
    \normalsize{} & \normalsize{\textbf{2005}} & \normalsize{\textbf{2006}} & \normalsize{\textbf{2007}}& \normalsize{\textbf{2008}}& \normalsize{\textbf{2009}} & \normalsize{\textbf{2010}}\\
    \hline
    \normalsize{\textbf{China}} & \normalsize{375} & \normalsize{430} & \normalsize{430}& \normalsize{545}& \normalsize{545} & \normalsize{545}\\

\hline
    \normalsize{\textbf{France}} & \normalsize{886} & \normalsize{1148} & \normalsize{1148}& \normalsize{1248}& \normalsize{1248} & \normalsize{1248}\\
    
    \hline    
     \normalsize{\textbf{Germany}} & \normalsize{765} & \normalsize{765} & \normalsize{765}& \normalsize{887}& \normalsize{937} & \normalsize{937}\\

    \hline
    \normalsize{\textbf{Italy}} & \normalsize{1217} & \normalsize{1217} & \normalsize{1217}& \normalsize{1231}& \normalsize{1231} & \normalsize{1245}\\
    
    \hline
    \normalsize{\textbf{Japan}} & \normalsize{957} & \normalsize{957} & \normalsize{957}& \normalsize{957}& \normalsize{957} & \normalsize{970}\\
    
    \hline
    \normalsize{\textbf{Netherlands}} & \normalsize{1005} & \normalsize{1005} & \normalsize{1005}& \normalsize{1020}& \normalsize{942} & \normalsize{942}\\    
    
     \hline
\normalsize{\textbf{United Kingdom}} & \normalsize{1267} & \normalsize{1267} & \normalsize{1267}& \normalsize{1350}& \normalsize{1160} & \normalsize{1045}\\
    \hline
    \normalsize{\textbf{United States}} & \normalsize{1160} & \normalsize{1160} & \normalsize{1160}& \normalsize{1245}& \normalsize{1315} & \normalsize{1315}\\
    \hline
	  \end{tabular}
		}}}\quad
    \subfigure[Diagram showing prominently answers (trend outliers) for task 3 on data  \cite{datasetTask3Web} \label{task3-outputs-pic}]{\includegraphics[width=9.5cm]{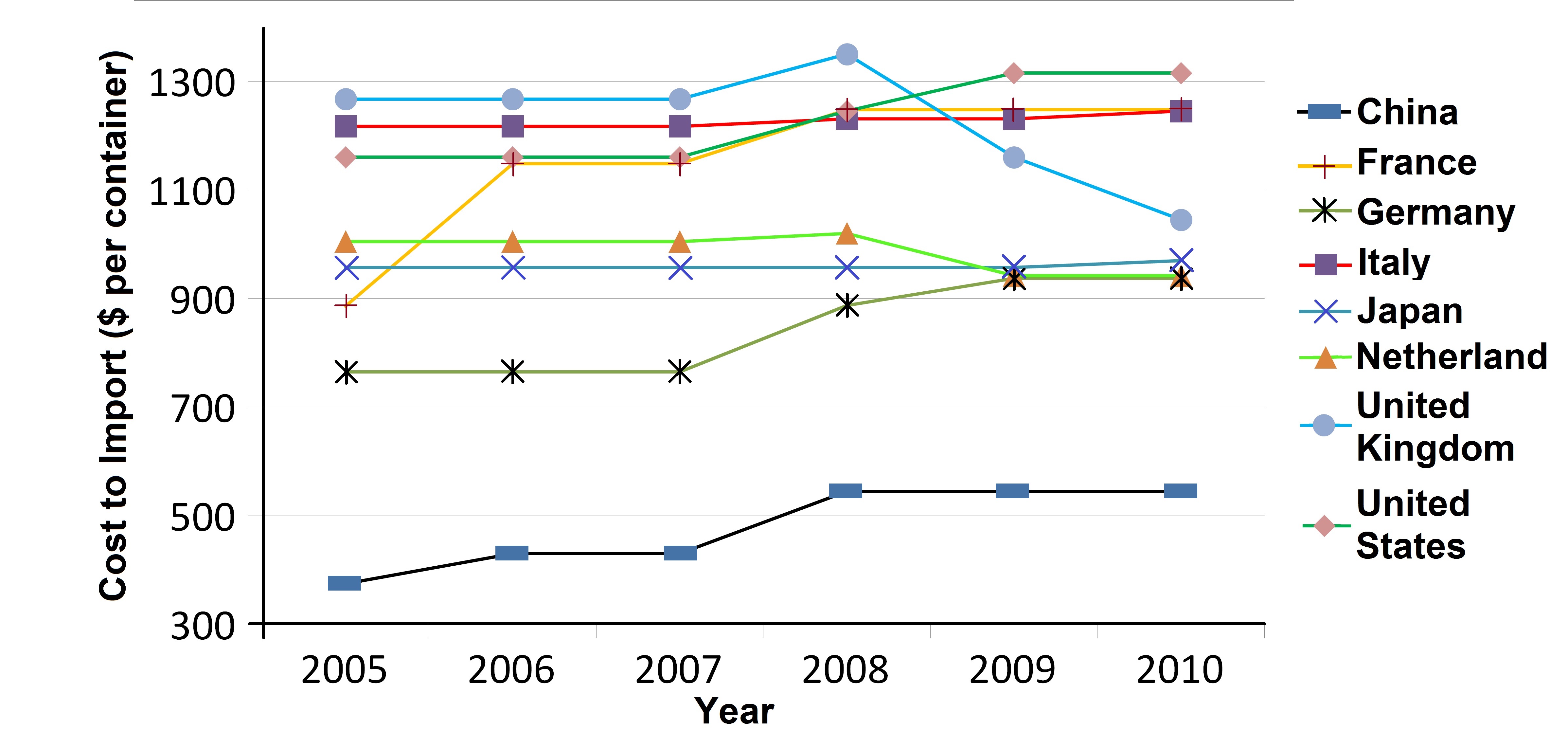}}\quad
    }
          \vspace{-5pt}
  \caption{Experimental task 3 using data set \cite{datasetTask3Web}: fragment of the data set and a visual solution that shows prominently the answers to the task.}
  \label{main figure label}
\end{figure*}

\begin{figure*}
  \centering
  \mbox{
    \subtable[Export and import values for China (top country in urban population) in billion US\$ in 2000--2010 in data set \cite{datasetTask3Web} \label{task4-data-rows-fig}]
	{\raisebox{7mm}{
		\resizebox{0.8\columnwidth}{!}{%
		\begin{tabular}[b]{|p{1.1cm}|M{1.0cm}|M{1.0cm}|M{1.0cm}|M{1.0cm}|}
    \hline
    \normalsize{} & \normalsize{\textbf{2000}} & \normalsize{\textbf{2001}} & \normalsize{\textbf{2002}}& \normalsize{\textbf{2003}}\\
    \hline
    \normalsize{\textbf{Export}} & \normalsize{279.56} & \normalsize{299.41} & \normalsize{365.40}& \normalsize{485.00}\\
    \hline    
     \normalsize{\textbf{Import}} & \normalsize{250.69} & \normalsize{271.33} & \normalsize{328.01}& \normalsize{448.92}\\
     \hline

    \normalsize{} & \normalsize{\textbf{2004}} & \normalsize{\textbf{2005}} & \normalsize{\textbf{2006}}& \normalsize{\textbf{2007}}\\
    \hline
    \normalsize{\textbf{Export}} & \normalsize{655.83} & \normalsize{836.90} & \normalsize{1061.68}& \normalsize{1342.21}\\
    \hline    
     \normalsize{\textbf{Import}} & \normalsize{606.54} & \normalsize{712.09} & \normalsize{852.77}& \normalsize{1034.73}\\
     \hline

    \normalsize{} & \normalsize{\textbf{2008}} & \normalsize{\textbf{2009}} & \normalsize{\textbf{2010}}& \normalsize{\textbf{}}\\
    \hline
    \normalsize{\textbf{Export}} & \normalsize{1581.71}& \normalsize{1333.30}& \normalsize{1752.10} & \normalsize{}\\

    \hline    
     \normalsize{\textbf{Import}} & \normalsize{1232.84}& \normalsize{1113.20}& \normalsize{1520.33}  & \normalsize{}\\
     \hline

	  \end{tabular}
		}}}\quad
    \subfigure[Line diagram showing  prominently answers (import/export trends) for the top country in urban population in 2000-2010, for task 4 on data \cite{datasetTask3Web} \label{task4-line diagram-pic}]{\includegraphics[width= 8.5cm]{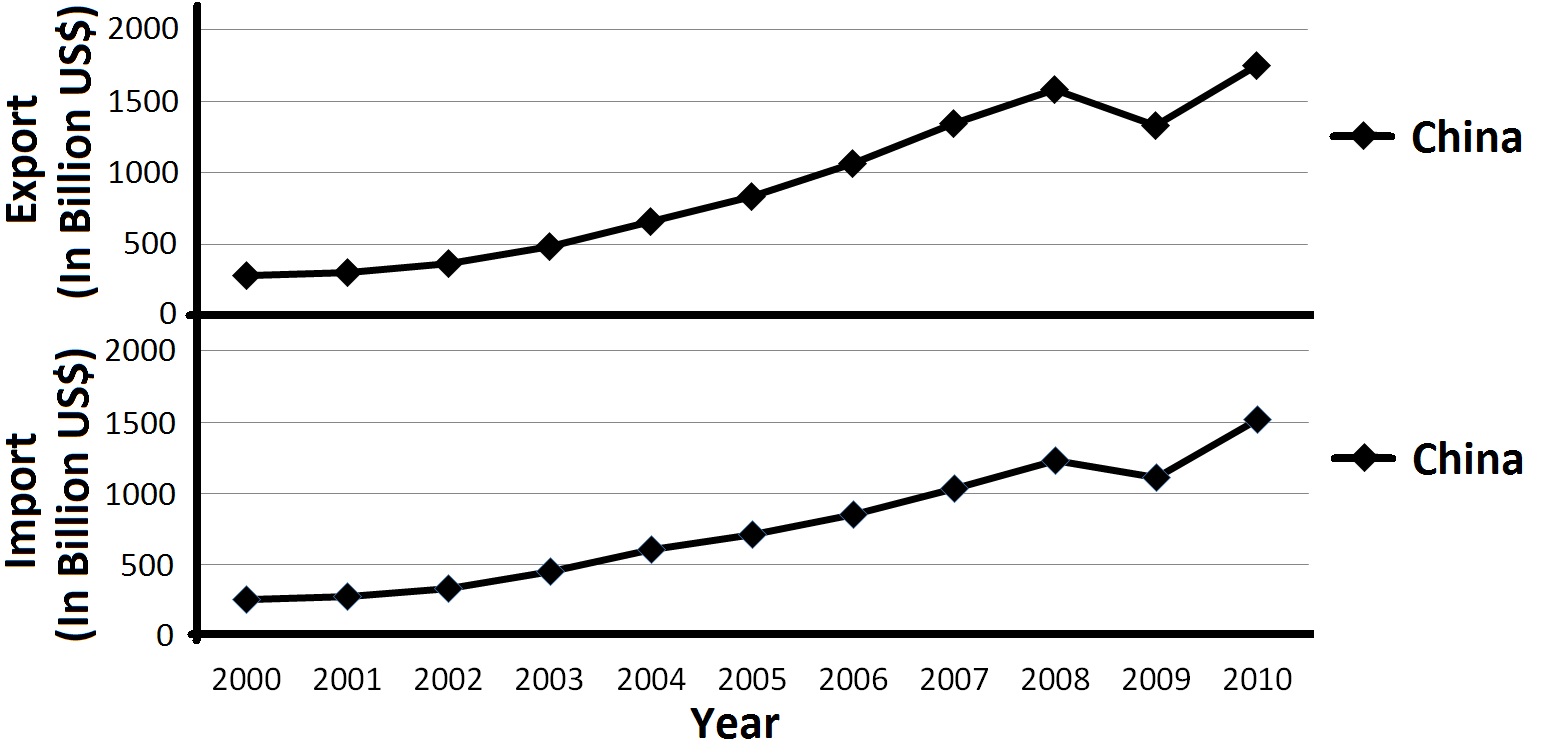}}\quad
    }
           \vspace{-5pt}
  \caption{Experimental task 4 using data set \cite{datasetTask3Web}: fragment of the data set and a visual solution that shows prominently the answers to the task.}
      \vspace{-10pt}
  \label{main figure label}
\end{figure*}

To evaluate DataSlicer's recommendation performance, we
conducted a set of controlled experiments. The results were evaluated in
terms of the measures of (see \cite{HealeyE12,HealeyS12}) {\em participant speed,} understood as  the average number of data-specification steps taken
to find a correct visualization for the task, as well as of {\em result accuracy,} understood as the degree to which the participant's solution is close to the correct solution. 
(In our experiments, the correct solutions were determined as part of the experimental setup.) Following the experiments,
each participant completed a questionnaire to capture their perception of:
(1) the difficulty of the assigned task, (2) 
the
correctness of their solution, (3) 
the correctness
of the system's solution, and (4) the overall usefulness of DataSlicer.

Both the statistical and questionnaire results were positive. Specifically, the results suggest that 
DataSlicer provides technically correct visualizations and, perhaps more 
importantly, rapidly directs participants to a correct visualization, 
potentially improving their performance over time. 

Due to the page limit, some of the discussions in this section are omitted. All of the omitted information can be found in the full version \cite{fullversion} of this paper.

\subsection{The Procedure}
 
We conducted four sets of experiments involving 48 human participants, with 12 
participants randomly assigned to each of the four  separate groups. The participants were graduate students ranging in age from 21 to 34, with 31 males and
17 females, each with  normal or corrected to normal
vision. 

Each of the 48 participants was first trained to work with our choice of front-end visualization software,
and was then given a task to complete. The tasks focused on common
data-analytics concepts of finding outliers and general data trends.  


After the initial training, each participant was asked to complete
their assigned task without using DataSlicer. The resulting log files
were analyzed for comparison with DataSlicer's recommended 
``correct'' visualization. Next, the participants used DataSlicer to find 
additional solutions for the same task, on the same data set. We then 
compared the accuracy and speed for the participants' 
task completion with and without access to DataSlicer. The participants concluded 
their session by providing feedback 
via a questionnaire (see \cite{fullversion}). 

The data sets used in the experiments are summarized in Table \ref{dataset-table}, and the experimental results are given in Table \ref{exp-results-simple}. Please note that the data sets (Table \ref{dataset-table}) were small in size. Still, we found (Table \ref{exp-results-simple}) that our human
participants had difficulty completing the assigned tasks even on
these small data sets. Presumably, increasing the number of
observations would further degrade the users' unassisted performance.




\subsection{The Tasks}

\definecolor{tbllightblue}{rgb}{0.7294,0.8471,0.8980}

\begin{table}[!t]
\centering
\setlength{\belowrulesep}{0pt}
\setlength{\aboverulesep}{0pt}
\renewcommand{\arraystretch}{1.4}
\begin{tabular}{l c p{1.75in}}
\toprule
  \rowcolor[gray]{0.7}
  \multicolumn{1}{c}{\sffamily Data set} &
  \multicolumn{1}{c}{\sffamily Observations} &
  \multicolumn{1}{c}{\sffamily Attributes} \\
\midrule
  \rowcolor[gray]{0.95}
  {\sffamily Earthquakes \cite{datasetTask1Web}} & {\sffamily 8289} &
  {\sffamily 17 attributes:
  Time,
  Date \& Time,
  Longitude,
  Latitude,
  Depth,
  Magnitude,
  Magnitude type,
  Nst,
  Gap,
  Dmin,
  Rms,
  Net,
  ID,
  Updated,
  Place,
  Type,
  Occurrences}\\
\midrule
  \rowcolor{tbllightblue}
  {\sffamily Baseball \cite{datasetTask2Web}} & {\sffamily 495} &
  {\sffamily 10 attributes:
  Name,
  Position,
  Type,
  AB,
  Age,
  BA,
  BB,
  G,
  H,
  RK}\\
\midrule
  \rowcolor[gray]{0.95}
  {\sffamily Economic \cite{datasetTask3Web}} & {\sffamily 2376} &
  {\sffamily 11 attributes:
  Country Name,
  Date,
  Exports,
  Imports,
  Cost to Import,
  Health Expenditure per Capita,
  Urban Population,
  Latitude,
  Longitude,
  Population Total,
  Health Expenditure Total}\\
\bottomrule
\end{tabular}
\vspace{5pt}
\caption[]{The data sets, number of observations, and number and names of attributes
  for the four experimental tasks.}
\label{dataset-table}
\end{table}

Each participant was asked to perform one of the four different tasks,
both with and without assistance from DataSlicer: (1) locating spatial
outliers in an earthquakes data set \cite{datasetTask1Web}; (2) locating data outliers in 
a baseball data set \cite{datasetTask2Web}; (3) locating outlier patterns and trends in 
an economic data set \cite{datasetTask3Web}; and (4) recognizing the general trends in the (same) data set \cite{datasetTask3Web}. The experiments were designed to cover common 
analytical tasks performed across a wide range of data domains; the tasks and data sets used in the experiments are as provided by \cite{jones2014communicating}. 


The expert sequences for each task were generated and validated as part of the experimental setup. The experts' log files 
were retrieved from the front-end visualization tool, parsed, and integrated 
into DataSlicer as discussed in Section~\ref{construc-sec}. 

\vspace{10pt}

\newcommand{\specialcell}[2][c]{%
  \begin{tabular}[#1]{@{}c@{}}#2\end{tabular}}

\begin{table*}[!t]
\centering
\setlength{\belowrulesep}{0pt}
\setlength{\aboverulesep}{0pt}
\setlength{\tabcolsep}{4pt}
\renewcommand{\arraystretch}{1.4}
%
%
\begin{minipage}{1.0\linewidth}
	\begin{minipage}{1.0\linewidth}
  \centering
  
  \begin{tabular}{cc}
  
  \begin{tabular}{c|c|c|c}
  \toprule
    \rowcolor[gray]{0.7}
    \multicolumn{4}{c}{\bfseries\sffamily Performance-Improvement Ratios for Users over Tasks 1-4}\\
  \midrule[0.2pt]
    \rowcolor[gray]{0.7}
    &\multicolumn{1}{c|}{\specialcell{\sffamily Minimum\\[-4pt]}} &
    \multicolumn{1}{c|}{\specialcell{\sffamily Maximum\\[-4pt]}} &
    \multicolumn{1}{c}{\specialcell{\sffamily Average\\[-4pt]}}\\

  \midrule
    \rowcolor[gray]{0.95}
    {\sffamily Accuracy-Improvement Ratios} &
    1.84 & 16.9& 5.09\\
%
%
    \rowcolor{tbllightblue}
    {\sffamily Speed-Improvement Ratios} &
    3.19 & 8.45 & 6.34\\
  \bottomrule
  \end{tabular} &
  \begin{tabular}{c|c|c|c|c}
  \toprule
    \rowcolor[gray]{0.7}
    \multicolumn{5}{c}{\bfseries\sffamily Average ($\mu$) User Speed per Task, in Visualization Steps Taken to Solve Task}\\
  \midrule[0.2pt]
    \rowcolor[gray]{0.7}{\bf Average speed (in visualization steps)} &
    \multicolumn{1}{c|}{\specialcell{\sffamily Task 1\\[-2pt] \itshape\sffamily (earthquake)}} &
    \multicolumn{1}{c|}{\specialcell{\sffamily Task 2\\[-2pt] \itshape\sffamily (baseball)}} &
    \multicolumn{1}{c|}{\specialcell{\sffamily Task 3\\[-2pt] \itshape\sffamily (economic)}}&
    \multicolumn{1}{c}{\specialcell{\sffamily Task 4\\[-2pt] \itshape\sffamily (economic)}}\\
    \rowcolor[gray]{0.95}
    {\sffamily Without DataSlicer} &
    17.1 & 16.9& 16.0 & 12.75\\
%
%
    \rowcolor{tbllightblue}
    {\sffamily With DataSlicer} &
    3 & 2 & 2 & 4\\
  \bottomrule
  \end{tabular}
    \vspace{-10pt}

   \\
  & \\
   {\sffamily\small (a)} & {\sffamily\small (b)}
  \end{tabular}
\end{minipage}

  \centering
  \begin{tabular}{c|c c|c c c c|c|c}
  \toprule
%
%
  \midrule[0.01pt]
    \rowcolor[gray]{0.7} {\bf Average ($\mu$) user }  &
    \multicolumn{2}{c|}{\sffamily Task 1} &
    \multicolumn{4}{c|}{\sffamily Task 2} &
    \multicolumn{1}{c|}{\sffamily Task 3} &
    \multicolumn{1}{c}{\sffamily Task 4}\\
    \rowcolor[gray]{0.7} {\bf accuracy per task} &
    \multicolumn{1}{c}{\specialcell{\itshape\sffamily  ``occurrence'' (\# \\[-2pt]\sffamily outliers found)}} &
    \multicolumn{1}{c|}{\specialcell{\itshape\sffamily  ``magnitude'' (\# \\[-2pt]\sffamily outliers found)}} &
    \multicolumn{1}{c}{\specialcell{\itshape\sffamily ``position'' (\# \\[-2pt]\sffamily attributes found)}} &
    \multicolumn{1}{c}{\specialcell{\itshape\sffamily ``position'' (\# \\[-2pt]\sffamily outliers found)}} &
    \multicolumn{1}{c}{\specialcell{\itshape\sffamily ``type'' (\# \\[-2pt]\sffamily attributes found)}} &
    \multicolumn{1}{c|}{\specialcell{\itshape\sffamily ``type'' (\# \\[-2pt]\sffamily outliers found)}} &
    \multicolumn{1}{c|}{\specialcell{\itshape\sffamily (\# outliers \\[-2pt] \sffamily in trends)}} &
    \multicolumn{1}{c}{\specialcell{\itshape\sffamily Correct visualization\\[-2pt] \sffamily achieved (\% cases) }}\\
    \rowcolor[gray]{0.95}
    {\sffamily Without DataSlicer} &
    4.9 & 9.9 & 0.58 & 0.17 & 0.42 & 0.92 & 0.5 & 50\\
%
%
    \rowcolor{tbllightblue}
    {\sffamily With DataSlicer} &
    83 & 30 & 2 & 1& 1 & 3 & 2 & 92\\
  \bottomrule
  \end{tabular}\\[4pt]
  {\sffamily\small (c)}
\end{minipage}\\[12pt]
%
%
  \vspace{-10pt}

\caption{Experimental results: (a) Performance improvements, reported for Accuracy as (With DataSlicer)/(Without DataSlicer) ratios, and for Speed as (Without DataSlicer)/(With DataSlicer) ratios; (b) Average speed values across the tasks; (c) Average user-accuracy values across the tasks.} 
%
%
\label{exp-results-simple}
\vspace{-10pt}
\end{table*}


\smallskip
\noindent
\textbf{Task 1: Spatial Outliers.} This task used an earthquakes 
data set \cite{datasetTask1Web} containing the location of 8,289 earthquakes with magnitude 6
or greater throughout the world, from 1900 to 2013
(Table~\ref{dataset-table}). The participants were asked to find \textit{places} (locations) on the map
that contain earthquakes with either: (1) outlier \textit{magnitudes;}
or (2) outlier \textit{number of occurrences}. (The definitions of outliers, via inter-quartile ranges, are ``as expected'' and can be found in \cite{fullversion}.) 

\smallskip
\noindent
\textbf{Task 2: Local Data Outliers.} This task used a baseball data set \cite{datasetTask2Web} 
containing information on 45 baseball players from the 2012 Major
League Baseball season (Table~\ref{dataset-table}). The participants were asked to find the data points for players that were outliers based on a
specific \textit{position} or \textit{type}. E.g., a
participant could look for outlier players at the shortstop position
by finding all shortstop players, then search for outliers
within that subgroup. If a data point contained any attribute that was an
outlier relative to the other players {\em in the subgroup} (hence the name ``{\em local} data outliers''), then that player  
would be reported as an outlier. (The definitions of outlier values, via inter-quartile ranges, are ``as expected'' and can be found in \cite{fullversion}.) 

\smallskip
\noindent
\textbf{Task 3: Outliers in Economic Patterns.} This task used a World Bank
indicators data set \cite{datasetTask3Web} containing 11 economic, health, and population
attributes for 216 countries 
for the
years 2000--2010 (Table~\ref{dataset-table}). 
The participants were asked to identify the top eight countries in terms
of average \textit{exports}, then determine which of these countries
displayed an outlier pattern in terms of \textit{export} statistics
over the given years. Outliers are identified by differences in
the direction of the slope of their trend lines versus the overall
norm for a given attribute. 

\smallskip
\noindent
\textbf{Task 4: General Economic Patterns.}
This task used the same data set \cite{datasetTask3Web} as Task 3. 
The participants were asked to identify a visualization that showed
the similarities and dissimilarities between the \textit{export} and \textit{import}
trends for the top country in the \textit{urban population} category over
the years 2000 to 2010.

\subsection{Expert Solutions}

We now discuss the steps that were used by experts  to solve tasks 3--4. (Due to the page limit, 
expert solutions for tasks 1--2 can be found in the full version \cite{fullversion} of the paper; Fig. \ref{task1-secondstage-pic} and Fig.~\ref{task2-outputs-pic} show the respective visualizations obtained by expert users to present the answers to the tasks.) 


%
%
%

\smallskip
\noindent
\textbf{Task 3.} Identifying \textit{export} pattern outliers in the
World Bank indicators data set \cite{datasetTask3Web} involved two stages. First, the top eight countries
in terms of average \textit{exports} were filtered by setting a lower
\textit{export} bound to include only eight countries. Next, a line-graph visualization of each country's \textit{exports} over the years
2000 to 2010 was generated. The countries whose trend lines deviated in
slope from the norm (\textit{i.e.,} the trend lines that did not follow the
ascending or descending pattern of the norm) were deemed to be
outliers (Fig. \ref{task3-outputs-pic}). 


\smallskip
\noindent
\textbf{Task 4.}
Recognizing general patterns in \textit{import} and \textit{export}
data for the top \textit{urban population} country in the World Bank indicators data set \cite{datasetTask3Web} involved two stages. First, the top \textit{urban population}
country in 2000--2010 was identified by setting a lower bound on
\textit{urban population} as a filter. Next, a line diagram was
generated on \textit{imports} and \textit{exports} over these years. The resulting visualization contains the top country's trends
for both \textit{imports} and \textit{exports} (Fig. \ref{task4-line diagram-pic}). 




\subsection{The Results}

The average results for accuracy (either the number of solutions found or the indicator of whether the single correct solution was found) and for speed
(the number of query steps performed), both without and with assistance
from DataSlicer, are detailed in Table~\ref{exp-results-simple}. Based on the average values in Table~\ref{exp-results-simple}, 
the accuracy of user solution for all tasks is at least 1.84 times better with DataSlicer than without DataSlicer, with the average of 5.09. 
Moreover, the speed in obtaining final visualization is at least 3 times better with DataSlicer than without DataSlicer, with the average of 6.34. 

We used Welch's analysis of variance (ANOVA) \cite{Welch51} to search for 
significant differences between the participant performance with and without
assistance from DataSlicer. Based on this analysis, we determined that in each of the four tasks,  
the participants were in statistically significant ways both faster\footnote{Increased speed here means that fewer data-specification operations were required with DataSlicer than without, to identify a correct
visualization.} and more accurate\footnote{Better accuracy here means that more outliers were located with DataSlicer than without, and general trends were located with DataSlicer but not without.}  with help from DataSlicer than without the help. (Due to the page limit, 
the report on the detailed statistics is omitted from this paper; the report can be found in  
the full version of the paper \cite{fullversion}.) 

Based on these results, we conclude that DataSlicer allows participants
to find statistically significantly more outliers and trends, in significantly fewer data-specification 
steps, than  unaided exploration. The tasks 
assigned to the participants include spatial outliers, local outliers, trend outliers, and general trends, which represent common analytic tasks on
real data.  Thus, the improved accuracy and speed in our experiments suggest 
better accuracy and speed 
for real-world data analysis.

The questionnaire results (see \cite{fullversion})
were also positive. On a scale of 1
to 7, with 1 being lowest and 7 highest, the participants rated the
usefulness of DataSlicer as 5.44, on average, and the accuracy of DataSlicer
as 5.94, on average. The participants were more confident about
their answers with DataSlicer than without (5.88 versus 5.46, on average).


\section{Conclusions}
\label{sec-conlcusions}
Searching for outlier data elements, data patterns, and
  trends are common and critical tasks during visual
analytics. The value of visualizations is in their offering the ability to present data in ways
that leverage a user's domain expertise, knowledge of context, and
ability to manage ambiguity that fully automated systems cannot. %
At the same time, users are often overwhelmed by the sheer volume of data (even in small data sets such as that \cite{datasetTask1Web} of experimental task 1 in Section \ref{experim-results-sec}), which may prevent 
them from understanding even basic properties of their data sets. This 
becomes particularly important in situations where the data set is 
large.

In our experiments with four task types designed to be representative of real-time exploration and discovery, DataSlicer significantly improved both the accuracy and speed for
identifying spatial outliers, data outliers, outlier patterns,
and general trends. The system quickly predicted what a
participant was searching for based on their initial operations, then presented recommendations that allowed the
participants to transform the data, leading them to 
identification of the desired solutions.

%
%
Although our data sets were moderate in size, the human
participants had difficulty completing the assigned tasks on the data. Presumably, increasing the size of data would further degrade their performance, and therefore
strengthen the value of using DataSlicer. As discussed in Section \ref{put-together-sec}, our predictive sequence
comparisons are relatively insensitive to data-set size, depending most
directly on the number of expert sequences to match against. In the
scenarios that we have tested, larger data sets would lead to more
target observations (e.g., outliers identified), but not to more steps
required to find the targets. In this way, we address an important
goal of scalability: With predictions based on user-generated  sequences, the prediction cost is based on the number of
sequences and sequence length, and not on data-set size.

We have run separate preliminary experiments with a ``recommendation'' data-slice graph involving only regular-user sequences from our original experiments with task 1 of Section \ref{experim-results-sec}.  The outcomes, discussed in \cite{fullversion}, 
 were far from satisfactory, as no graph nodes were of significant help to users in their solving the task with DataSlicer. 
This confirms the intuition that such tasks are very difficult to solve for users that are not experts in their fields, therefore reinforcing the desirability of 
constructing data-slice graphs using expert sequences. It remains to be seen if recommendation graphs can be useful tools for simpler tasks  
or with significantly larger user bases. 

%
%
%



\bibliographystyle{plain}
\bibliography{submitted101915}

\begin{thebibliography}{10}

\bibitem{fullversion}
F.~Alborzi, S.~Chaudhuri, and et~al.
\newblock Data{S}licer: Enabling data selection for visual data exploration.
\newblock Technical Report TR-2015-8, NCSU, 2015.
\newblock {\tt http://www.csc.ncsu.edu/research/tech/reports.php}.

\bibitem{cetintemel2013query}
U.~Cetintemel, M.~Cherniack, J.~DeBrabant, Y.~Diao, K.~Dimitriadou, A.~Kalinin,
  O.~Papaemmanouil, and S.~Zdonik.
\newblock Query steering for interactive data exploration.
\newblock In {\em CIDR}, 2013.

\bibitem{chatzopoulou2009query}
G.~Chatzopoulou, M.~Eirinaki, and N.~Polyzotis.
\newblock Query recommendations for interactive database exploration.
\newblock {\em Scientific and Statistical Database Management}, pages 3--18,
  2009.

\bibitem{dimitriadou2014explore}
K.~Dimitriadou, O.~Papaemmanouil, and Y.~Diao.
\newblock Explore-by-example: An automatic query steering framework for
  interactive data exploration.
\newblock In {\em ACM SIGMOD}, pages 517--528, 2014.

\bibitem{datasetTask1Web}
{Dropbox}.
\newblock
  \url{https://www.dropbox.com/s/hye4pi82wcwsrrp/CDWT_ch7_Earthquakes.xlsx}.
\newblock Accessed in February 2015.

\bibitem{datasetTask2Web}
{Dropbox}.
\newblock
  \url{https://www.dropbox.com/s/eflow6vsmgbulk5/CDWT_ch5_2012NYStats.xlsx}.
\newblock Accessed in February 2015.

\bibitem{datasetTask3Web}
{Dropbox}.
\newblock
  \url{https://www.dropbox.com/s/jym44gtqni2qddf/Sample%20-%20World%20Bank%20Indicators.xlsx}.
\newblock Accessed in February 2015.

\bibitem{drosou2013ymaldb}
M.~Drosou and E.~Pitoura.
\newblock Ymaldb: exploring relational databases via result-driven
  recommendations.
\newblock {\em VLDBJ}, 22:849--874, 2013.

\bibitem{fan2011interactive}
J.~Fan, G.~Li, and L.~Zhou.
\newblock Interactive {SQL} query suggestion: Making databases user-friendly.
\newblock In {\em ICDE}, pages 351--362, 2011.

\bibitem{GotzW09}
D.~Gotz and Z.~Wen.
\newblock Behavior-driven visualization recommendation.
\newblock {\em Proc. Intl' Conf. Intelligent User Interfaces}, pages 315--324,
  2009.

\bibitem{GrammelTS10}
L.~Grammel, M.~Tory, and M~Storey.
\newblock How information visualization novices construct visualizations.
\newblock {\em IEEE Trans. Visualiz. and Comp. Graph.}, 16(6):943--952, 2010.

\bibitem{HealeyD12}
C.~G. Healey and B.~M. Dennis.
\newblock Interest driven navigation in visualization.
\newblock {\em IEEE Trans. Visualiz. and Comp. Graph.}, 18:1744--1756, 2012.

\bibitem{HealeyE12}
C.~G. Healey and J.~T. Enns.
\newblock Attention and visual memory in visualization and computer graphics.
\newblock {\em {IEEE} Trans. Vis. Comput. Graph.}, 18(7):1170--1188, 2012.

\bibitem{HealeyS12}
C.~G. Healey and A.~P. Sawant.
\newblock On the limits of resolution and visual angle in visualization.
\newblock {\em {ACM} {T}rans. {A}pplied {P}erception}, 9(4):20, 2012.

\bibitem{idreos2015overview}
S.~Idreos, O.~Papaemmanouil, and S.~Chaudhuri.
\newblock Overview of data exploration techniques.
\newblock In {\em ACM SIGMOD}, pages 277--281, 2015.

\bibitem{jones2014communicating}
Ben Jones.
\newblock {\em Communicating Data with Tableau}.
\newblock O'Reilly Media, 2014.

\bibitem{key2012vizdeck}
A.~Key, B.~Howe, D.~Perry, and C.~Aragon.
\newblock Vizdeck: self-organizing dashboards for visual analytics.
\newblock In {\em ACM SIGMOD}, 2012.

\bibitem{khoussainova2010snipsuggest}
N.~Khoussainova, Y.~Kwon, M.~Balazinska, and D.~Suciu.
\newblock Snip{S}uggest: Context-aware autocompletion for {SQL}.
\newblock {\em PVLDB}, 4(1):22--33, 2010.

\bibitem{LivnyRBCDLMW97}
M.~Livny, R.~Ramakrishnan, K.~S. Beyer, G.~Chen, D.~Donjerkovic, S.~Lawande,
  J.~Myllymaki, and R.~K. Wenger.
\newblock {DEV}ise: Integrated querying and visual exploration of large
  datasets.
\newblock In {\em ACM SIGMOD}, pages 517--520, 1997.

\bibitem{MackinlayHS07}
J.~D. Mackinlay, P.~Hanrahan, and C.~Stolte.
\newblock Show {M}e: Automatic presentation for visual analysis.
\newblock {\em {IEEE} Trans. Vis. Comput. Graph.}, 13(6):1137--1144, 2007.

\bibitem{neophytou2012astroshelf}
P.~Neophytou, R.~Gheorghiu, R.~Hachey, T.~Luciani, D.~Bao, A.~Labrinidis,
  E.~Marai, and P.~Chrysanthis.
\newblock Astro{S}helf: understanding the universe through scalable navigation
  of a galaxy of annotations.
\newblock In {\em ACM SIGMOD}, pages 713--716, 2012.

\bibitem{NiuZRS12}
F.~Niu, C.~Zhang, C.~Re, and J.~Shavlik.
\newblock Deep{D}ive: {W}eb-scale knowledge-base construction using statistical
  learning and inference.
\newblock In {\em Proc. Intl' Wkshp Searching Integrating New Web Data
  Sources}, 2012.

\bibitem{parameswaran2013seedb}
A.~Parameswaran, N.~Polyzotis, and H.~Garcia-Molina.
\newblock See{DB}: Visualizing database queries efficiently.
\newblock {\em PVLDB}, 7(4):325--328, 2013.

\bibitem{sellam2013meet}
T.~Sellam and M.~Kersten.
\newblock Meet {C}harles, big data query advisor.
\newblock In {\em CIDR}, 2013.

\bibitem{ShinWWSZR15}
J.~Shin, S.~Wu, F.~Wang, C.~De Sa, C.~Zhang, and C.~R{\'{e}}.
\newblock Incremental knowledge base construction using {D}eep{D}ive.
\newblock {\em {PVLDB}}, 8, 2015.

\bibitem{StolperPG14}
C.~Stolper, A.~Perer, and D.~Gotz.
\newblock Progressive visual analytics: User-driven visual exploration of
  in-progress analytics.
\newblock {\em {IEEE} Trans. Vis. Comput. Graph.}, 20(12):1653--1662, 2014.

\bibitem{PolarisDiss}
C.~Stolte.
\newblock {\em Query, analysis, and visualization of multidimensional
  databases}.
\newblock PhD thesis, Stanford University, 2003.

\bibitem{StolteTH08}
C.~Stolte, D.~Tang, and P.~Hanrahan.
\newblock Polaris: a system for query, analysis, and visualization of
  multidimensional databases.
\newblock {\em Comm. {ACM}}, 51(11):75--84, 2008.

\bibitem{vartak2014seedb}
M.~Vartak, S.~Madden, A.~Parameswaran, and N.~Polyzotis.
\newblock See{DB}: automatically generating query visualizations.
\newblock {\em PVLDB}, 7(13), 2014.

\bibitem{wasayqueriosity}
A.~Wasay, M.~Athanassoulis, and S.~Idreos.
\newblock Queriosity: Automated data exploration.
\newblock In {\em Proc. {IEEE} International Congress on {B}ig {D}ata}, 2015.

\bibitem{Welch51}
B.~L. Welch.
\newblock On the comparison of several mean values: An alternative approach.
\newblock {\em Biometrika}, 3/4:330--336, 1951.

\end{thebibliography}


\end{document}